\documentclass[aps,twocolumn,pra,showpacs,amsmath,amssymb,showkeys,10pt]{revtex4-2}
\usepackage{graphicx}
\usepackage{epstopdf}
\pdfoutput=1
\usepackage{braket}
\usepackage{hyperref}
\usepackage{longtable}
\newcommand{\tabincell}[2]{\begin{tabular}{@{}#1@{}}#2\end{tabular}}
\begin{document}

	\title{Investigating the quantum discord dynamics with a bipartite split of the multiqubit system in the correlated photon--matter model}
	
	\author{Hui-hui Miao}
	\email[Correspondence to: Vorobyovy Gory 1, Moscow, 119991, Russia. Email address: ]{hhmiao@cs.msu.ru}
	\affiliation{Faculty of Computational Mathematics and Cybernetics, Lomonosov Moscow State University, Vorobyovy Gory 1, Moscow, 119991, Russia}

	\date{\today}

	\begin{abstract}
	In this paper, we try to study the quantum discord dynamics in a complex correlated photon--matter model, which is modified from the Tavis--Cummings--Hubbard model --- a common cavity quantum electrodynamics model. The target model consists of two hydrogen atoms. A neutral hydrogen molecule can be obtained through an association reaction and disintegrated through dissociation reaction. The formation and breaking of covalent bond is accompanied by the creation and annihilation of phonon. Compared with previous efforts, studying the quantum discord dynamics of this complicated system is more challenging than it was for the simple quantum system, which consisted of a single two-level atom. For convenience, we adopt a bipartite split of the multiqubit system and the two-qubit von Neumann projective measurement on the observed subsystem. We attempt to examine the dissipative dynamics in open quantum system in addition to the unitary evolution of closed quantum system. We are dedicated to identifying the regularity of quantum correlation as the basis for future research on more complex quantum systems, specifically including the impacts of nuclei tunneling effect, covalent bond formation strength, and dissipation intensities of photon (phonon) on quantum discord.
	\end{abstract}

	%\pacs{03.65.Yz, 42.50.Lc, 42.50.-p, 42.50.Pq}
	\keywords{quantum discord, quantum correlation, bipartite system, finite-dimensional QED, neutron hydrogen molecule.}

	\maketitle

	\section{Introduction}
	\label{sec:Introduction}
	
	Information on the state of a quantum system is known as quantum information. It is the fundamental object of study in quantum information theory \cite{Vedral2006, Nielsen2010, Masahito2006}, and it can be changed using quantum information processing (QIP) techniques. Classical information is measured using Shannon entropy, while the quantum mechanical analogue is von Neumann entropy \cite{Bengtsson2006, Nielsen2010, Zachos2007}. Given a statistical ensemble of quantum mechanical systems with the density matrix 
$\rho$ , it is given by $S(\rho)=-tr(\rho ln\rho)$ \cite{Nielsen2010}. Many of the same entropy measures in classical information theory can also be generalized to the quantum case, such as Holevo entropy \cite{Holevo1973} and the conditional quantum entropy \cite{Cerf1997, Cerf1999, HorodeckiOppenheim2005}. In comparison to the traditional method, QIP, which is closely related to quantum entanglement \cite{Einstein1935, Horodecki2009}, provides a greater variety of information manipulation techniques. Due to its uniqueness, quantum entanglement has been viewed as a crucial resource for QIP, including quantum computation \cite{Nielsen2010}, quantum teleportation \cite{Bennett1993, Horodecki1995}, superdense coding \cite{Bennett1992}, remote state preparation \cite{Pati2000}, quantum cryptography \cite{Ekert1991, Gisin2002}, and many more. However, quantum entanglement is simply a specific kind of quantum correlation that combines classical and quantum components. Ollivier et al. hypothesized and explained in detail the quantum discord \cite{Ollivier2001, Zurek2003, Henderson2001, Vedral2003, Dakic2010} --- a novel candidate of quantum correlation. In mathematical terms, quantum discord is defined in terms of the quantum mutual information. More specifically, quantum discord is the difference between two expressions which both represent the mutual information in the classical case. These two expressions are
	\begin{equation}
		\label{eq:ClassicalExpression1}
		I\left(A:B\right)=H\left(A\right)+H\left(B\right)-H\left(AB\right)
	\end{equation}
	\begin{equation}
		\label{eq:ClassicalExpression2}
		J\left(A:B\right)=H\left(A\right)-H\left(A|B\right)
	\end{equation}
where, in the classical case, $H\left(A\right)$ (or $H\left(B\right)$) is the information entropy, $H\left(AB\right)$ the joint entropy and $H\left(A|B\right)$ the conditional entropy. $H\left(A|B\right)$ has following form
	\begin{equation}
		\label{eq:ConditionalEntropy}
		H\left(A|B\right)=H\left(AB\right)-H\left(B\right)
	\end{equation}
Thus, the Eqs. \eqref{eq:ClassicalExpression1} and \eqref{eq:ClassicalExpression2} yield identical results. In the nonclassical case, the quantum physics analogy for the three terms $H\left(A\right)$, $H\left(AB\right)$ and $H\left(A|B\right)$ are used --- $S\left(\mathcal{A}\right)$ the von Neumann entropy, $S\left(\mathcal{AB}\right)$ the joint quantum entropy and $S\left(\mathcal{A}|\mathcal{B}\right)$ a quantum generalization of conditional entropy, respectively. Here, $\mathcal{AB}\equiv\rho_{\mathcal{AB}}$, $\mathcal{A}\equiv\rho_{\mathcal{A}}=Tr_{\mathcal{B}}(\rho_{\mathcal{AB}})$ and $\mathcal{B}\equiv\rho_{\mathcal{B}}=Tr_{\mathcal{A}}(\rho_{\mathcal{AB}})$, where $\rho_{\mathcal{AB}}$ is density matrix of entire quantum system and $\rho_{\mathcal{A}}$ ($\rho_{\mathcal{B}}$) is reduced density matrix. Thus, the quantum analogues of Eqs. \eqref{eq:ClassicalExpression1} and \eqref{eq:ClassicalExpression2} are defined as follows
	\begin{equation}
		\label{eq:QuantumExpression1}
		\mathcal{I}\left(\mathcal{A}:\mathcal{B}\right)=S\left(\mathcal{A}\right)+S\left(\mathcal{B}\right)-S\left(\mathcal{AB}\right)
	\end{equation}
	\begin{equation}
		\label{eq:QuantumExpression2}
		\mathcal{J}\left(\mathcal{A}:\mathcal{B}\right)=S\left(\mathcal{A}\right)-S\left(\mathcal{A}|\mathcal{B}\right)
	\end{equation}
Now the difference between the two expressions defines the quantum discord. Quantum discord was recently introduced into many research fields \cite{WangJieci2010, Fanchini2010, Chakrabarty2011, HuYaoHua2012, XieMeiQiu2013, FanKaiMing2013, LiRuiQi2014, Aldoshin2014, Guo2015, JiaZhihAhn2020, Radhakrishnan2020, Brown2021, MiaoLi2023}. Additionally, other quantum correlation measurement methods similar to quantum discord are proposed, such as, quantum dissonance \cite{Modi2010}, quantum deficit\cite{Rajagopal2002, Horodecki2005, Devetak2005} and quantumness of correlations \cite{Devi2008}, which are temporarily not considered in this paper.

	The quantum electrodynamics (QED) model is a fundamental contribution to this paper. We introduce $\eta$, which is defined as follows 
	\begin{equation}
		\label{eq:Eta}
		\eta=max\left(\frac{g}{\hbar\omega_c},\frac{g}{\hbar\omega_a}\right)
	\end{equation}
where $\hbar=h/2\pi$ --- the reduced Planck constant or Dirac constant, $h$ --- the Planck constant, $g$ is the coupling strength, $\omega_c$ stands for cavity frequency and $\omega_a$ for transition frequency. According to Eq. \eqref{eq:Eta}, the QED model can be divided into three types: ultrastrong-coupling (USC), deep strong coupling (DSC) and strong coupling (SC). The USC \cite{Rabi1936, Rabi1937, Dicke1954, Hopfield1958, Haroche2013, Gu2017, Kockum2019, Frisk2019, Forn-Diaz2019} occurs when coupling strength $g$ becomes comparable to the atomic ($\omega_a$) or cavity ($\omega_c$) frequencies. More specifically, the USC regime occurs when $\eta$ is within the range $\left[0.1, 1\right)$. The DSC \cite{Casanova2010} is a common term used to describe the regime $\eta\geq 1$. The more straightforward SC model --- Jaynes--Cummings model (JCM) \cite{Jaynes1963} occurs when $\eta<0.1$, and depicts the dynamics of a two-level atom in an optical cavity, interacting with a single-mode field inside it. Its generalization --- the Tavis--Cummings model (TCM) \cite{Tavis1968} depicts the dynamics of a collection of $N$ two-level atoms in an optical cavity. The Jaynes--Cummings--Hubbard model (JCHM) and Tavis--Cummings--Hubbard model (TCHM) \cite{Angelakis2007} are generalizations of the JCM and TCM to multiple cavities coupled by an optical fibre. Some efforts have been carried out in the area of cavity QED models \cite{Lee1999, Andre2002, Tanamoto2012, Poltl2013, Hansom2014, Ferretti2016, Prasad2018, GuoLijuan2019, Quach2020, Ozhigov2020, OzhigovYI2020, Rose2021, WeiHuanhuan2021, Smith2021, Dull2021, Ozhigov2021, Ozhigov2022, Kulagin2022, Miao2023, MiaoOzhigov2023}.
	
	The target model, called the association--dissociation model of neutral hydrogen molecule, which is a modified version of the TCHM and has been studied in our earlier work \cite{Miao2023}, involves the creation and annihilation of phonon. Phonon is the quasiparticle that represents simple harmonic oscillation. In contrast to previous articles, which use one-qubit von Neumann projective measurement to achieve quantum discord in two-qubit quantum systems, in this paper, we focus on the seven-qubit system and try to adopt the two-qubit measurement on the observed subsystem. Both closed and open quantum systems are considered in this paper. The dissipative dynamics can be obtained in open system by solving the quantum master equation (QME). Consideration is also given to the impacts of nuclei tunneling effect, covalent bond formation strength and dissipation intensities of photon (phonon) on quantum discord.
	
	This paper is organized as follows. After explaining in detail the quantum discord with a bipartite split of the multiqubit system in Sec. \ref{sec:Discord}, we introduce the target model in Sec. \ref{sec:Target}. Then we do some simulations and get the results in Sec. \ref{sec:Results}. Some brief comments on our results and extension to future work in Sec. \ref{sec:ConcluFuture} close out the paper. Some technique details are shown in Appendices \ref{appx:TwoQubitvonNeumann}, \ref{appx:SecondQuantization}, \ref{appx:Operators}, \ref{appx:QME} and \ref{appx:Method}. List of abbreviations and notations used in this paper is put in Appendix \ref{appx:AbbreviationsNotations}.
	
	\section{Quantum discord}
	\label{sec:Discord}
	
	\begin{figure}
		\begin{center}
		\includegraphics[width=0.5\textwidth]{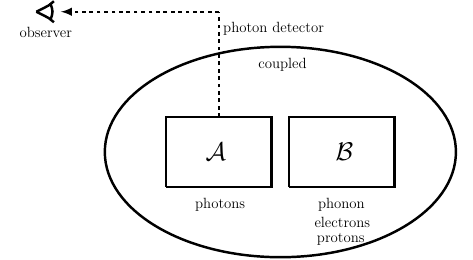}
		\end{center}
		\caption{(online color) {\it The coupling system of apparatus and substance.} The multiqubit system is separated into two subsystems: the observed subsystem $\mathcal{A}$ and the substance subsystem $\mathcal{B}$. In this paper, the substance subsystem consists of phonon, electrons and atomic nuclei (protons), and the part observed by the apparatus is photons (here observer uses a photon detector to capture photons). Subsystems $\mathcal{A}$ and $\mathcal{B}$ are coupled together into a whole by the field in the optical cavity.}
		\label{fig:CouplingSystem}
	\end{figure}

	As shown in Fig. \ref{fig:CouplingSystem}, for a bipartite system, we set the observed subsystem as $\mathcal{A}$ (photon) and the unobserved subsystem as $\mathcal{B}$ (phonon, electrons, protons). Subsystems are coupled together into a whole by the field in the optical cavity. Quantum discord is a measure of nonclassical correlations between two subsystems. Moreover, it serves as a gauge for the quantumness of correlations. The quantum discord for target model is defined as follows
	\begin{equation}
		\label{eq:QuantumDiscord}
		\mathcal{D}(\mathcal{B}:\mathcal{A})=\mathcal{I}(\mathcal{B}:\mathcal{A})-\mathcal{J}(\mathcal{B}:\mathcal{A})
	\end{equation}
	where $0\leq \mathcal{D}(\mathcal{B}:\mathcal{A})<\mathcal{I}(\mathcal{B}:\mathcal{A}),\ \mathcal{D}(\mathcal{B}:\mathcal{A})\leq S(\mathcal{A})$. $\mathcal{I}(\mathcal{B}:\mathcal{A})$ is identical to $\mathcal{I}(\mathcal{A}:\mathcal{B})$, which is defined in Eq. \eqref{eq:QuantumExpression1}. The notation $\mathcal{J}(\mathcal{B}:\mathcal{A})$ represents the part of the correlations that can be attributed to classical correlations and varies in dependence on the chosen eigenbasis; therefore, in order for the quantum discord to reflect the purely nonclassical correlations independently of basis, it is necessary that $\mathcal{J}(\mathcal{B}:\mathcal{A})$ first be maximized over the set of all possible projective measurements onto the eigenbasis \cite{Datta2009, Zurek2007}
	\begin{equation}
		\label{eq:ClassicalCorrelationBA}
		\begin{aligned}
			\mathcal{J}(\mathcal{B}:\mathcal{A})&=\underset{\{\Pi^{\mathcal{A}}_k\}}{max}[S(\mathcal{B})-S(\mathcal{B}|\mathcal{A})]\\
			&=\underset{\{\Pi^{\mathcal{A}}_k\}}{max}[S(\mathcal{B})-\sum_kp_kS(\rho_k)]\\
			&=S(\mathcal{B})-\underset{\{\Pi^{\mathcal{A}}_k\}}{min}\sum_kp_kS(\rho_k)
		\end{aligned}
	\end{equation}
	where $\{\Pi_k^{\mathcal{A}}\}$ is a complete set of projectors for the two-qubit von Neumann projective measurement of subsystem $\mathcal{A}$. The von Neumann projective measurement is carried out on the subsystem $\mathcal{A}$ with dimension $4$ (two qubits). The technique details of the two-qubit von Neumann projective measurement used in this paper can be found in Appendix \ref{appx:TwoQubitvonNeumann}, where we defined in detail the von Neumann measurement basis $\{|b_k\rangle\}$ and the set of operators $\{\Pi_k^{\mathcal{A}}\}$. The maximization in $\mathcal{J}(\mathcal{B}:\mathcal{A})$ represents the most information obtained about the subsystem $\mathcal{B}$ that can be gathered from the perfect measurement $\{\Pi_k^{\mathcal{A}}\}$. The quantum discord can be demonstrated to be nonzero for states with quantum correlation and zero for states with just classical correlation. After constructing the projection operators of two-qubit observed subsystem $\mathcal{A}$, we can measure it and get $\rho_k$, which is defined as follows
	\begin{equation}
		\label{eq:RhokBA}
		\rho_k=Tr_{\mathcal{A}}[(\Pi_k^{\mathcal{A}}\otimes I_{\mathcal{B}})\rho_{\mathcal{AB}}(\Pi_k^{\mathcal{A}}\otimes I_{\mathcal{B}})^{\dag}]/p_k
	\end{equation}
	where
	\begin{equation}
		\label{eq:pkBA}
		p_k=Tr[(\Pi_k^{\mathcal{A}}\otimes I_{\mathcal{B}})\rho_{\mathcal{AB}}(\Pi_k^{\mathcal{A}}\otimes I_{\mathcal{B}})^{\dag}]
	\end{equation}
	$\rho_k$ is the state of the subsystem $\mathcal{B}$ after a measurement of subsystem $\mathcal{A}$ leading to an outcome $k$ with a probability $p_k$. 
	
	In most cases, quantum discord is asymmetrical in the sense that $\mathcal{D}(\mathcal{B}:\mathcal{A})$ can differ from $\mathcal{D}(\mathcal{A}:\mathcal{B})$ \cite{Dakic2010, Dillenschneider2008}, because $\mathcal{J}(\mathcal{B}:\mathcal{A})$ is not identical to $\mathcal{J}(\mathcal{A}:\mathcal{B})$, which is defined as follows
	\begin{equation}
		\label{eq:ClassicalCorrelationAB}
		\begin{aligned}
			\mathcal{J}(\mathcal{A}:\mathcal{B})&=\underset{\{\Pi^{\mathcal{B}}_{k'}\}}{max}[S(\mathcal{A})-S(\mathcal{A}|\mathcal{B})]\\
			&=\underset{\{\Pi^{\mathcal{B}}_{k'}\}}{max}[S(\mathcal{A})-\sum_{k'}p_{k'}S(\rho_{k'})]\\
			&=S(\mathcal{A})-\underset{\{\Pi^{\mathcal{B}}_{k'}\}}{min}\sum_{k'}p_{k'}S(\rho_{k'})
		\end{aligned}
	\end{equation}
	where $\{\Pi_{k'}^{\mathcal{B}}\}$ is a complete set of projectors for the five-qubit von Neumann projective measurement of subsystem $\mathcal{B}$. Then, we can measure it and get $\rho_{k'}$, which is defined as follows
	\begin{equation}
		\label{eq:RhokAB}
		\rho_{k'}=Tr_{\mathcal{B}}[(I_{\mathcal{A}}\otimes\Pi_{k'}^{\mathcal{B}})\rho_{\mathcal{AB}}(I_{\mathcal{A}}\otimes\Pi_{k'}^{\mathcal{B}})^{\dag}]/p_{k'}
	\end{equation}
	where
	\begin{equation}
		\label{eq:pkAB}
		p_{k'}=Tr[(I_{\mathcal{A}}\otimes\Pi_{k'}^{\mathcal{B}})\rho_{\mathcal{AB}}(I_{\mathcal{A}}\otimes\Pi_{k'}^{\mathcal{B}})^{\dag}]
	\end{equation}
	$\rho_{k'}$ is the state of the subsystem $\mathcal{A}$ after a measurement of subsystem $\mathcal{B}$ leading to an outcome $k'$ with a probability $p_{k'}$. 
	
	The greater quantum discord $\mathcal{D}(\mathcal{B}:\mathcal{A})$, the greater the minimum loss of information in the subsystem $\mathcal{B}$ after measurement of subsystem $\mathcal{A}$. This is the physical meaning of quantum discord. In other words, the greater the minimum disturbance caused to subsystem $\mathcal{B}$ after measuring the subsystem $\mathcal{A}$, the stronger the correlation between $\mathcal{B}$ and $\mathcal{A}$.
	
	\section{Target model}
	\label{sec:Target}
	
	\begin{figure}
		\begin{center}
		\includegraphics[width=0.45\textwidth]{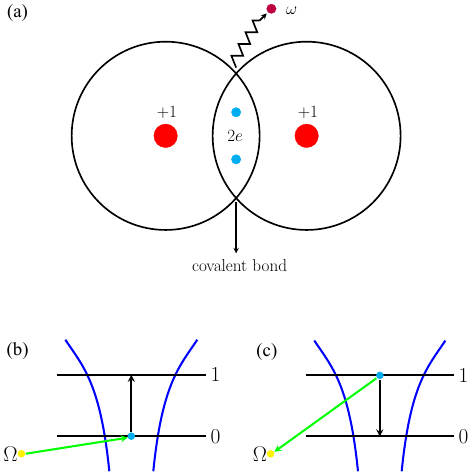}
		\end{center}
		\caption{(online color) {\it The schematic diagram of the target model.} The process of molecular hydrogen formation between two hydrogen atoms, accompanied by covalent bond formation and release of phonon, is shown in panel (a). Panels (b) and (c) show two kinds of atom--field interactions (excitation and relaxation) corresponding to photonic mode $\Omega$. In our target model, each interaction with field can be considered as a separate JCM, and the target model consists of many JCM. In the panels (a), (b) and (c), proton, electron, photon and phonon are seen as red, cyan, yellow and purple dot, respectively.}
		\label{fig:TargetModel}
	\end{figure}
	
	\begin{figure*}
		\begin{center}
		\includegraphics[width=1.\textwidth]{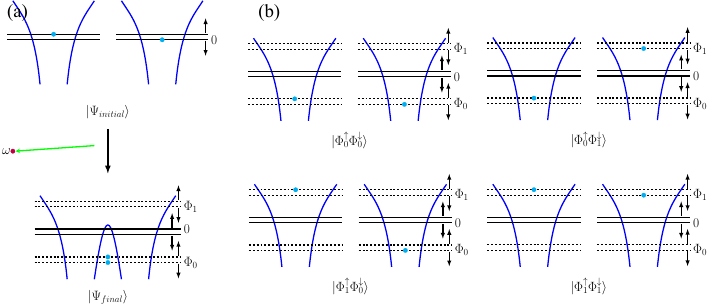}
		\end{center}
		\caption{(online color) {\it The initial and final states.} When nuclei are in the same cavity in the model depicted in panel (a), electrons can jump between orbitals $\Phi_1$ and $\Phi_0$ after hybridization of atomic orbitals to molecular orbitals, covalent bond are formed at the same time. Or they can be constrained to orbital $0$ after de-hybridization of molecular orbitals to atomic orbitals, covalent bond are broken at the same time. In order to make a covalent bond, a phonon must be released, and in order to break a covalent bond, a phonon must be absorbed. In panel (b), according to Eqs. \eqref{eq:MolState}, $|\Psi_{initial}\rangle$ can be decomposed into the sum of four states $|\Phi_0^{\uparrow}\Phi_0^{\downarrow}\rangle$, $|\Phi_0^{\uparrow}\Phi_1^{\downarrow}\rangle$, $|\Phi_1^{\uparrow}\Phi_0^{\downarrow}\rangle$ and $|\Phi_1^{\uparrow}\Phi_1^{\downarrow}\rangle$. In the panels (a) and (b), electron and phonon are seen as cyan and yellow dot, respectively. The figure is cited and modified from \cite{Miao2023}.}
		\label{fig:InitialState}
	\end{figure*}
	
	The target model, featuring a covalent bond and a phonon, is presented in Fig. \ref{fig:TargetModel} (a), which shows that two hydrogen atoms combine to form a neutral hydrogen molecule and release a phonon. Here two independent hydrogen atoms are close together, the atomic orbitals hybridize into molecular orbitals, and the two electrons located in the molecular orbitals form a covalent bond. The hybridization progress of atomic orbitals is shown in Fig. \ref{fig:InitialState} (a). Atomic orbital $0$ is hybridized into the molecular orbitals: ground $\Phi_0$ and excited $\Phi_1$. In a closed system, electrons oscillate in these two molecular orbitals, but in a dissipative system, as the photons escape, the electrons eventually fall to the ground orbital. Panels (b) and (c) of Fig. \ref{fig:TargetModel} represent respectively the excitation and de-excitation processes of hydrogen atom --- a typical two-level atom. During the excitation of an atom, an electron jumps from the ground state to the excited state and absorbs a photon. On the contrary, during the de-excitation process, the electron falls from the unstable excited state to the ground state and releases a photon.
	
	The system composed of two hydrogen atoms is called the association-dissociation model of neutral hydrogen molecule, and its detailed description is put in \cite{Miao2023}. Our target model is modified from this model: compared with original model, in the modified version, we ignore the excitation and relaxation of electrons in atomic orbitals and default to electrons always being on atomic orbital $0$, and study the evolution and quantum discord of the system when only the excitation and relaxation of electrons in molecular orbitals are retained. In this model, each energy level, both atomic and molecular, is split into two levels with the same energy (approximately the same, with accuracy to Stark splitting): spin up and spin down, which are indicated by the signs $\uparrow$ and $\downarrow$, respectively. To differentiate each level on the spin, we will add these marks that indicate the energy level. Now the levels will be twice as much, and for each level there must be no more than one electron according to Pauli exclusion principle \cite{Pauli1925}. Thus, photons that excite the electron will be of the same type as the chosen spin direction.
	
	In this paper, we adopt the second quantization (see Appendix \ref{appx:SecondQuantization}) for describing the quantum states. The Hilbert space of quantum states of the entire system has the following form
	\begin{equation}
		\label{eq:SpaceBondPhonon}
		|\Psi\rangle_{\mathcal{C}}=\underbrace{|p_1\rangle_{\Omega^{\uparrow}}|p_2\rangle_{\Omega^{\downarrow}}}_{\mathcal{A}}\underbrace{|m\rangle_{\omega}|l_1\rangle_{\Phi_1^{\uparrow}}|l_2\rangle_{\Phi_1^{\downarrow}}|L\rangle_{cb}|k\rangle_{n}}_{\mathcal{B}}
	\end{equation}
	where $p_1=0,\ 1$ is the number of molecular photons with mode $\Omega^{\uparrow}$, $p_2=0,\ 1$ is the number of molecular photons with mode $\Omega^{\downarrow}$, $m=0,\ 1$ is the number of phonons with mode $\omega$. Spin up and spin down are denoted by $\uparrow$ and $\downarrow$, respectively. $l_1,\ l_2$ describe orbital state: $l_1=1$ --- electron with spin $\uparrow$ in excited orbital $\Phi_1^{\uparrow}$, $l_1=0$ --- electron with spin $\uparrow$ in ground orbital $\Phi_0^{\uparrow}$; $l_2=1$ --- electron with spin $\downarrow$ in excited orbital $\Phi_1^{\downarrow}$, $l_2=0$ --- electron with spin $\downarrow$ in ground orbital $\Phi_0^{\downarrow}$. The states of the covalent bond are denoted by $|L\rangle_{cb}$: $L=0$ --- formation, $L=1$ --- breaking. The relative position between atomic nuclei is denoted by $|k\rangle_n$: $k=0$ --- atoms gathering together in one cavity, $k=1$ --- atoms scattering in different cavities. Here Hilbert space is divided into to subsystems: $|p_1\rangle_{\Omega^{\uparrow}}|p_2\rangle_{\Omega^{\downarrow}}$ is considered as the two-qubit observed subsystem $\mathcal{A}$ and $|m\rangle_{\omega}|l_1\rangle_{\Phi_1^{\uparrow}}|l_2\rangle_{\Phi_1^{\downarrow}}|L\rangle_{cb}|k\rangle_{n}$ is considered as the five-qubit substance subsystem $\mathcal{B}$.
	
	Before constructing the Hamiltonian, we first introduce rotating wave approximation (RWA) \cite{Wu2007}, which is taken into account
	\begin{equation}
		\label{eq:RWACondition}
		\frac{g}{\hbar\omega_a}\approx\frac{g}{\hbar\omega_c}\ll 1
	\end{equation}
Usually, for convenience, we assume that $\omega_a$ and $\omega_c$ are equal. In cavity QED models, Hamiltonian usually contain the terms
	\begin{equation}
		\label{eq:InteractionTerms}
		\left(a^{\dag}+a\right)\left(\sigma^{\dag}+\sigma\right)=a^{\dag}\sigma^{\dag}+a\sigma^{\dag}+a^{\dag}\sigma+a\sigma
	\end{equation}
where $a$ is photon annihilation operator, $a^{\dag}$ is photon creation operator --- hermitian conjugate operator of $a$, $\sigma$ is electron relaxation operator, and $\sigma^{\dag}$ is electron excitation operator --- hermitian conjugate operator of $\sigma$. RWA allows us to ignore the quickly oscillating terms $\sigma^{\dag}a^{\dag},\ \sigma a$ in a Hamiltonian, so we can change $\left(\sigma^{\dag}+\sigma\right)\left(a^{\dag}+a\right)$ to $\sigma^{\dag}a+\sigma a^{\dag}$. Thus, Hamiltonian of the target model has following form
	\begin{equation}
		\label{eq:HamilBondPhonon}
		\begin{aligned}
			H_{sys}&=\hbar\Omega^{\uparrow}a_{\Omega^{\uparrow}}^{\dag}a_{\Omega^{\uparrow}}+\hbar\Omega^{\downarrow}a_{\Omega^{\downarrow}}^{\dag}a_{\Omega^{\downarrow}}+\hbar\omega a_{\omega}^{\dag}a_{\omega}\\
			&+\hbar\Omega^{\uparrow}\sigma_{\Omega^{\uparrow}}^{\dag}\sigma_{\Omega^{\uparrow}}+\hbar\Omega^{\downarrow}\sigma_{\Omega^{\downarrow}}^{\dag}\sigma_{\Omega^{\downarrow}}+\hbar\omega\sigma_{\omega}^{\dag}\sigma_{\omega}\\
			&+g_{\Omega^{\uparrow}}\left(a_{\Omega^{\uparrow}}^{\dag}\sigma_{\Omega^{\uparrow}}+a_{\Omega^{\uparrow}}\sigma_{\Omega^{\uparrow}}^{\dag}\right)\sigma_{\omega}\sigma_{\omega}^{\dag}\\
			&+g_{\Omega^{\downarrow}}\left(a_{\Omega^{\downarrow}}^{\dag}\sigma_{\Omega^{\downarrow}}+a_{\Omega^{\downarrow}}\sigma_{\Omega^{\downarrow}}^{\dag}\right)\sigma_{\omega}\sigma_{\omega}^{\dag}\\
			&+g_{\omega}\left(a_{\omega}^{\dag}\sigma_{\omega}+a_{\omega}\sigma_{\omega}^{\dag}\right)+\zeta\left(\sigma_n^{\dag}\sigma_n+\sigma_n\sigma_n^{\dag}\right)
		\end{aligned}
	\end{equation}
	where $\zeta$ is the tunnelling strength, $a_{\Omega^{\uparrow}}$ and $a_{\Omega^{\uparrow}}^{\dag}$ are photon annihilation and creation operators for mode $\Omega_{\uparrow}$, $a_{\Omega^{\downarrow}}$ and $a_{\Omega^{\downarrow}}^{\dag}$ are photon annihilation and creation operators for mode $\Omega_{\downarrow}$, $a_{\omega}$ and $a_{\omega}^{\dag}$ are phonon annihilation and creation operators for mode $\omega$, $\sigma_{\Omega^{\uparrow,\downarrow}}$ and $\sigma_{\Omega^{\uparrow,\downarrow}}^{\dag}$ are the electron relaxation and excitation operators, $\sigma_{\omega}$ and $\sigma_{\omega}^{\dag}$ are the covalent bond's formation and break operators, $\sigma_n$ and $\sigma_n^{\dag}$ are the nuclei's tunneling operators. And $\sigma_{\omega}\sigma_{\omega}^{\dag}$ verifies that covalent bond is formed. The values of $\Omega^{\uparrow}$ and $\Omega^{\downarrow}$ are almost equal. How all operators used in Eq. \eqref{eq:HamilBondPhonon} act on quantum state will be shown in Appendix \ref{appx:Operators}.
	
	 The process of hybridization of atomic orbitals ($|0_1\rangle$ corresponds to one atom, $|0_2\rangle$ corresponds to another atom) into molecular orbitals ($|\Phi_0\rangle$ and $|\Phi_1\rangle$) is shown in Fig. \ref{fig:InitialState} (a). And $|\Phi_0\rangle,\ |\Phi_1\rangle$ have following forms
	\begin{subequations}
		\label{eq:MolState}
		\begin{align}
			|\Phi_0\rangle=\frac{1}{\sqrt{2}}\left(|0_1\rangle+|0_2\rangle\right)\label{eq:MolState0}\\
			|\Phi_1\rangle=\frac{1}{\sqrt{2}}\left(|0_1\rangle-|0_2\rangle\right)\label{eq:MolState1}
		\end{align}
	\end{subequations}
	We can obtain the expression of $|0_1\rangle$, $|0_2\rangle$ from Eq. \eqref{eq:MolState}
	\begin{subequations}
		\label{eq:AtomState}
		\begin{align}
			|0_1\rangle=\frac{1}{\sqrt{2}}\left(|\Phi_0\rangle+|\Phi_1\rangle\right)\label{eq:AtomState0}\\
			|0_2\rangle=\frac{1}{\sqrt{2}}\left(|\Phi_0\rangle-|\Phi_1\rangle\right)\label{eq:AtomState1}
		\end{align}
	\end{subequations}
	
		\begin{table}[!htpb]
        \centering
        \caption{List of quantum states.}	
		\label{Tab:States}
		\begin{tabular}{|l|l||l|l||l|l|}
			\hline
			Index & State & Index & State & Index & State \\
			\hline
			0 & $|0000000\rangle$ & 9 & $|0000111\rangle$ & 18 & $|0110000\rangle$ \\
			\hline
		 	1 & $|0100000\rangle$ & 10 & $|0001000\rangle$ & 19 & $|1010000\rangle$ \\
			\hline
		 	2 & $|1000000\rangle$ & 11 & $|0101000\rangle$ & 20 & $|1110000\rangle$ \\
			\hline
		 	3 & $|1100000\rangle$ & 12 & $|0001010\rangle$ & 21 & $|0010100\rangle$ \\
			\hline
		 	4 & $|0000010\rangle$ & 13 & $|0001011\rangle$ & 22 & $|1010100\rangle$ \\
			\hline
		 	5 & $|0000011\rangle$ & 14 & $|0001100\rangle$ & 23 & $|0011000\rangle$ \\
			\hline
		 	6 & $|0000100\rangle$ & 15 & $|0001110\rangle$ & 24 & $|0111000\rangle$ \\
			\hline
		 	7 & $|1000100\rangle$ & 16 & $|0001111\rangle$ & 25 & $|0011100\rangle$ \\
			\hline
		 	8 & $|0000110\rangle$ & 17 & $|0010000\rangle$ & & \\
			\hline
		\end{tabular}
	\end{table}
	
	Fig. \ref{fig:InitialState} (a) shows that the diatomic system evolves from an initial state (two independent hydrogen atoms and covalent bond is broken) to a final state (stable hydrogen molecule and covalent bond is formed). According to Eq. \eqref{eq:AtomState}, $|\Psi_{initial}\rangle$ can be expressed by adding the four components, which are represented on Fig. \ref{fig:InitialState} (b)
	\begin{equation}
		\label{eq:InitialDecomposed}
		\begin{aligned}
			|\Psi_{initial}\rangle&=|0_1^{\uparrow}\rangle|0_2^{\downarrow}\rangle\\
			&=\frac{1}{\sqrt{2}}\left(|\Phi_0^{\uparrow}\rangle+|\Phi_1^{\uparrow}\rangle\right)\frac{1}{\sqrt{2}}\left(|\Phi_0^{\downarrow}\rangle-|\Phi_1^{\downarrow}\rangle\right)\\
			&=\frac{1}{2}\left(|\Phi_0^{\uparrow}\Phi_0^{\downarrow}\rangle-|\Phi_0^{\uparrow}\Phi_1^{\downarrow}\rangle+|\Phi_1^{\uparrow}\Phi_0^{\downarrow}\rangle-|\Phi_1^{\uparrow}\Phi_1^{\downarrow}\rangle\right)
		\end{aligned}
	\end{equation}
	where
	\begin{subequations}
		\label{eq:PhiPhi}
		\begin{align}
			|\Phi_0^{\uparrow}\Phi_0^{\downarrow}\rangle=|0\rangle_{\omega^{\uparrow}}|0\rangle_{\omega^{\downarrow}}|0\rangle_{\omega}|0\rangle_{\Phi_1^{\uparrow}}|0\rangle_{\Phi_1^{\downarrow}}|1\rangle_{cb}|0\rangle_{n}\label{eq:Phi0Phi0}\\
			|\Phi_0^{\uparrow}\Phi_1^{\downarrow}\rangle=|0\rangle_{\omega^{\uparrow}}|0\rangle_{\omega^{\downarrow}}|0\rangle_{\omega}|0\rangle_{\Phi_1^{\uparrow}}|1\rangle_{\Phi_1^{\downarrow}}|1\rangle_{cb}|0\rangle_{n}\label{eq:Phi0Phi1}\\
			|\Phi_1^{\uparrow}\Phi_0^{\downarrow}\rangle=|0\rangle_{\omega^{\uparrow}}|0\rangle_{\omega^{\downarrow}}|0\rangle_{\omega}|1\rangle_{\Phi_1^{\uparrow}}|0\rangle_{\Phi_1^{\downarrow}}|1\rangle_{cb}|0\rangle_{n}\label{eq:Phi1Phi0}\\
			|\Phi_1^{\uparrow}\Phi_1^{\downarrow}\rangle=|0\rangle_{\omega^{\uparrow}}|0\rangle_{\omega^{\downarrow}}|0\rangle_{\omega}|1\rangle_{\Phi_1^{\uparrow}}|1\rangle_{\Phi_1^{\downarrow}}|1\rangle_{cb}|0\rangle_{n}\label{eq:Phi1Phi1}
		\end{align}
	\end{subequations}
	
	The final state $|\Psi_{final}\rangle=|\Phi_0^{\uparrow}\Phi_0^{\downarrow}\rangle$ represents that two electrons with different spins are situated on the molecular ground orbital, and all photons and phonon have escaped from the optical cavity. Thus, stable hydrogen molecule is obtained in open system. In the closed case, since escape of photons and phonon is not considered, the entire system will oscillate between two situations: two independent hydrogen atoms and a synthetic hydrogen molecule.
	
	The QME in the Markovian approximation for the density operator $\rho$ of the open system takes the following form
	\begin{equation}
		\label{eq:QME}
		i\hbar\dot{\rho}=\left[H_{sys},\rho\right]+iL(\rho)
	\end{equation}
	where $H_{sys}$ is system Hamiltonian, $\rho$ is density matrix and $L(\rho)$ is superoperator. $\left[H_{sys},\rho\right]=H_{sys}\rho-\rho H_{sys}$ is the commutator. $L(\rho)$ contains many jump operators, which describe the escape of particles from the optical cavity to the external environment, and some jump operators also describe the influx of particles into the optical cavity from the external environment. Some technique details about QME in Markovian open system are put in Appendix \ref{appx:QME}.
	
	\begin{figure*}
		\begin{center}
		\includegraphics[width=1.\textwidth]{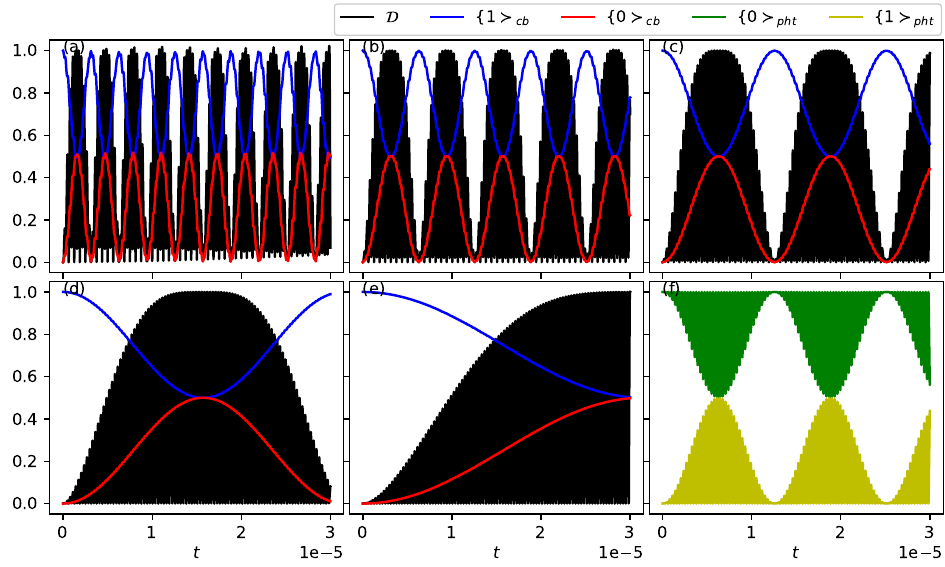}
		\end{center}
		\caption{(online color) {\it Quantum discord dynamics in closed system with tunneling effect.} Here dissipative rate $\gamma=0$, thus unitary evolution is obtained. We assumed that the values of strengths $g_{\Omega^{\uparrow}}$ and $g_{\Omega^{\downarrow}}$ are equal to $g$, and tunneling effect is allowed, thus $\zeta=g$. We study the effect of different values of $g_{\omega}$ from large to small on the unitary evolution and quantum discord dynamics: (a) $g_{\omega}=0.2g$, (b) $g_{\omega}=0.1g$, (c) $g_{\omega}=0.05g$, (d) $g_{\omega}=0.02g$, (e) $g_{\omega}=0.01g$. In addition, in order to compare the relationship between photon number and quantum discord dynamics, we take the change of photonic states ($\left\{0\succ_{pht}\right.$ and $\left\{1\succ_{pht}\right.$, which are defined in Eq. \eqref{eq:StatePhts}) when $g_{\omega}=0.05g$, shown in panel (f), and compare it with quantum discord dynamics, shown in panel (c).}
		\label{fig:ClosedSystemWithTunneling}
	\end{figure*}
	
	The second quantization expressions of quantum states involved in quantum evolution are shown in Tab. \ref{Tab:States}, where only 26 states are written, which is less than the total number of quantum states $2^7=128$, corresponding to system's seven qubits. The numerical method for solution $\rho(t)$ in Eq. \ref{eq:QME} is shown in Appendix \ref{appx:Method}. And in this Appendix, introduced is also the generator algorithm \cite{Miao2023}, which is dedicated to eliminate the extra unnecessary states for reducing the complexity. In other words, this method shows how we can compress a Hilbert space $\mathcal{C}$ of 128 states into a space $\mathcal{C}'$ of only 26 states. According to the new Hilbert space $\mathcal{C}'$, we can construct a simpler Hamiltonian $H_{sys}'$ that takes up less computer memory and costs less time.
	
	\section{Results}
	\label{sec:Results}

	\begin{figure*}
		\begin{center}
		\includegraphics[width=1.\textwidth]{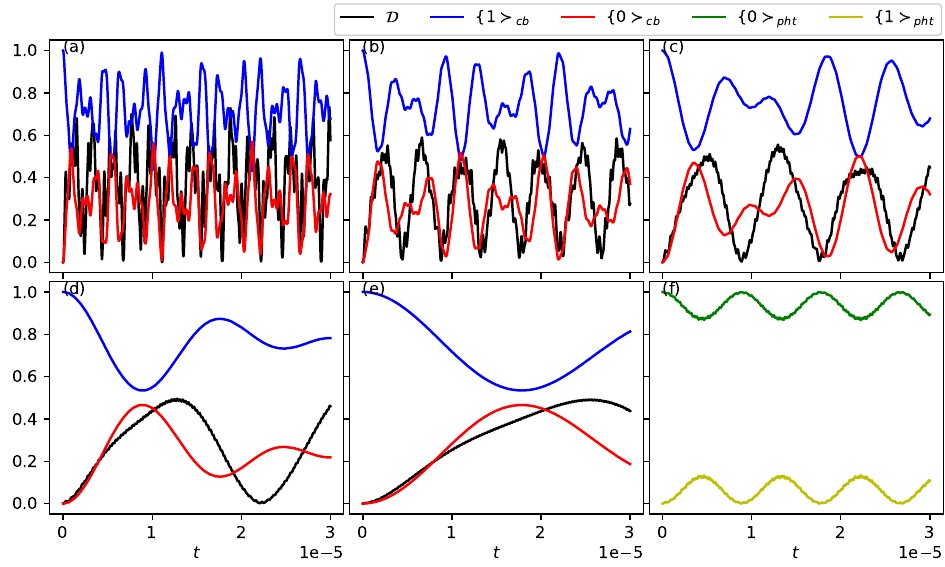}
		\end{center}
		\caption{(online color) {\it Quantum discord dynamics in closed system without tunneling effect.} Unitary evolution is also obtained. However, the tunneling effect is now forbidden, thus $\zeta=0$. In this case, we again study the effect of different values of $g_{\omega}$ on the unitary evolution and quantum discord dynamics: (a) $g_{\omega}=0.2g$, (b) $g_{\omega}=0.1g$, (c) $g_{\omega}=0.05g$, (d) $g_{\omega}=0.02g$, (e) $g_{\omega}=0.01g$, and compare the photonic states $\left\{0\succ_{pht}\right.$ and $\left\{1\succ_{pht}\right.$ with quantum discord dynamics when $g_{\omega}=0.05g$.}
		\label{fig:ClosedSystemWithoutTunneling}
	\end{figure*}
	
	\subsection{Closed system} 
	\label{subsec:ClosedSystem}
	
	First, initial state $|\Psi_{initial}\rangle$ describes that two hydrogen atoms are both in the same cavity and will move to different cavities due to tunneling effect. We assume $g=10^7$, and $g_{\Omega^{\uparrow}}=g_{\Omega^{\downarrow}}=\zeta=g,\ \gamma_{\Omega^{\uparrow}}=\gamma_{\Omega^{\downarrow}}=\gamma_{\omega}=0$. Now we define two states
	\begin{subequations}
		\label{eq:StateCBs}
		\begin{align}
			\left\{0\succ_{cb}\right.&=\sum_{p_1,p_2,m,l_1,l_2,k}c|p_1\rangle|p_2\rangle|m\rangle|l_1\rangle|l_2\rangle|0\rangle|k\rangle\label{eq:StateCB0}\\
			\left\{1\succ_{cb}\right.&=\sum_{p_1,p_2,m,l_1,l_2,k}c'|p_1\rangle|p_2\rangle|m\rangle|l_1\rangle|l_2\rangle|1\rangle|k\rangle\label{eq:StateCB1}
		\end{align}
	\end{subequations}
	where $c,\ c'$ are normalization factors, which depends on $p_1,\ p_2,\ m,\ l_1,\ l_2,\ k$. $\left\{0\succ_{cb}\right.$ --- covalent bond is formed, that is to say, system is in molecular state; $\left\{1\succ_{cb}\right.$ --- covalent bond is broken, and system is in diatomic state.
	
	\subsubsection{With tunneling effect} 
	\label{subsubsec:WithTunneling}
	
	\begin{figure}
		\begin{center}
		\includegraphics[width=0.45\textwidth]{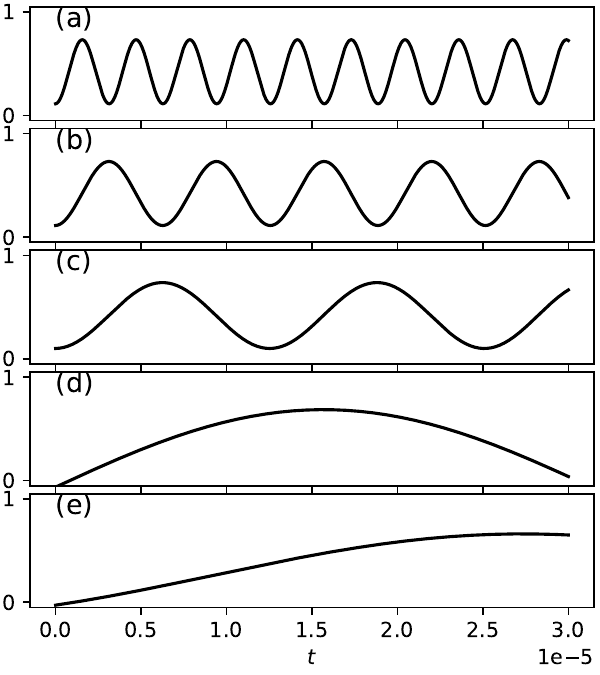}
		\end{center}
		\caption{(online color) {\it $sin$-like function fitting of quantum discord varies with $g_{\omega}$ in case of tunneling effect.} Different values of $g$ are represented from panel (a) to (e): (a) $g_{\omega}=0.2g$, (b) $g_{\omega}=0.1g$, (c) $g_{\omega}=0.05g$, (d) $g_{\omega}=0.02g$, (e) $g_{\omega}=0.01g$. $\zeta=g$.}
		\label{fig:FittingFig4}
	\end{figure}
	
	\begin{figure}
		\begin{center}
		\includegraphics[width=0.45\textwidth]{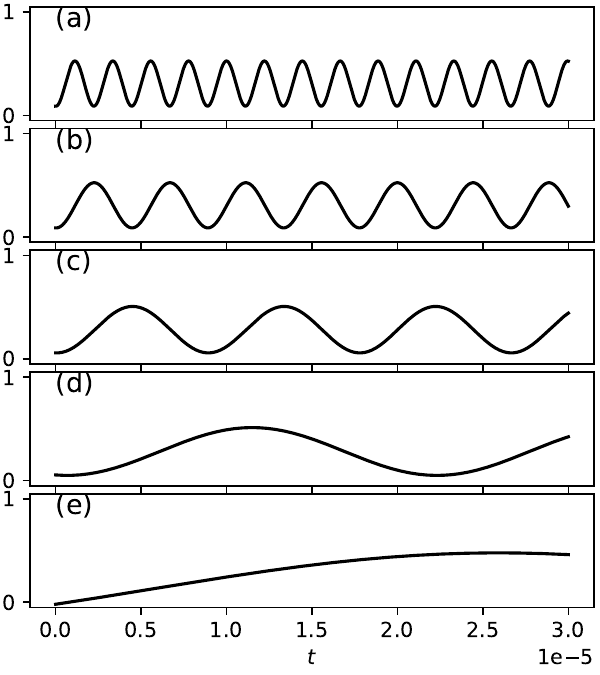}
		\end{center}
		\caption{(online color) {\it $sin$-like function fitting of quantum discord varies with $g_{\omega}$ without the tunneling effect.} Similar to Fig. \ref{fig:FittingFig4}, different values of $g$ are represented from panel (a) to (e): (a) $g_{\omega}=0.2g$, (b) $g_{\omega}=0.1g$, (c) $g_{\omega}=0.05g$, (d) $g_{\omega}=0.02g$, (e) $g_{\omega}=0.01g$. $\zeta=0$.}
		\label{fig:FittingFig5}
	\end{figure}
	
	\begin{figure}
		\begin{center}
		\includegraphics[width=0.5\textwidth]{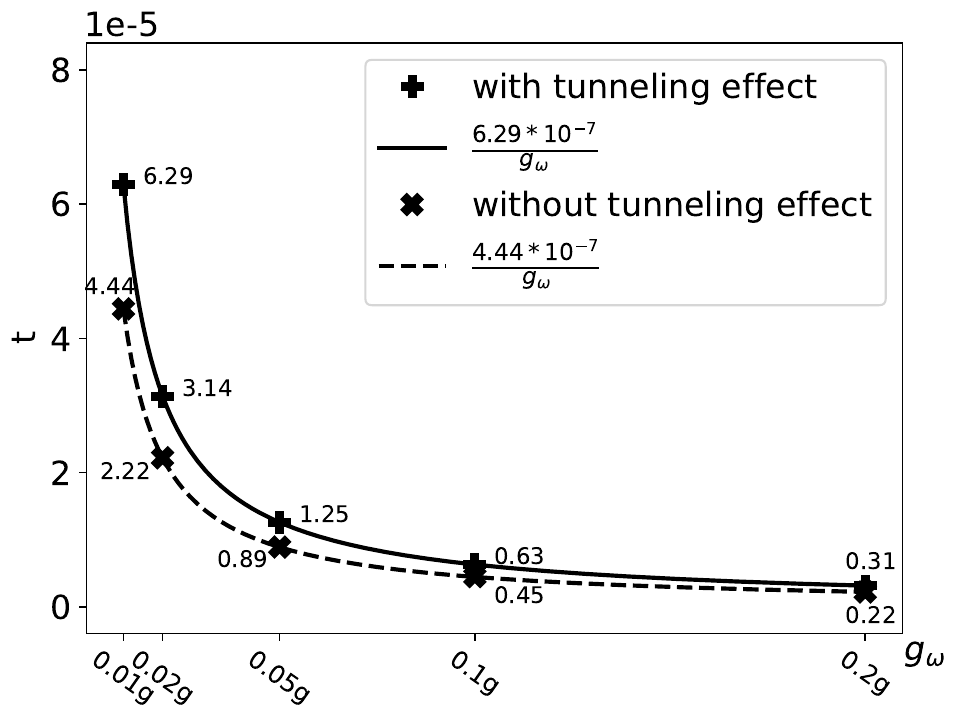}
		\end{center}
		\caption{(online color) {\it Period of the $sin$-like function fitting curves of quantum discord from Figs. \ref{fig:FittingFig4} and \ref{fig:FittingFig5}.} The bold markers "+" is corresponding to the case of tunneling effect and the bold markers "$\times$" is corresponding to the case without tunneling effect. Solid curve is inverse proportional relationship with form $\frac{6.29*10^{-7}}{g_{\omega}}$ and dashed curve is inverse proportional relationship with form $\frac{4.44*10^{-7}}{g_{\omega}}$, where $g_{\omega}$ is independent variable (abscissa). The value of abscissa from small to large is $0.01g,\ 0.02g,\ 0.05g,\ 0.1g,\ 0.2g.$}
		\label{fig:Comparison}
	\end{figure}
	
	From Fig. \ref{fig:ClosedSystemWithTunneling} (a) $\sim$ (e), we can find that $\left\{0\succ_{cb}\right.$ and $\left\{1\succ_{cb}\right.$ form synchronous oscillations during the quantum evolution process, and their curves are wavy, when $g_{\omega}$ is comparable to $g$. When the value of $g_{\omega}$ is getting smaller and smaller than $g$, the curves of $\left\{0\succ_{cb}\right.$ and $\left\{1\succ_{cb}\right.$ becomes smoother. The reason is that when two atoms are close together, they may either tunnel away under the action of strength $\zeta=g$ (the system stays in state $\left\{1\succ_{cb}\right.$) or they may form molecule through orbital hybridization under the action of strength $g_{\omega}$ (the system stays in state $\left\{0\succ_{cb}\right.$). When these two strengths are close in magnitude, the confrontation between $\left\{0\succ_{cb}\right.$ and $\left\{1\succ_{cb}\right.$ is fierce, and when one is much smaller than the other, the confrontation becomes negligible.
	
	Now we focus on quantum discord. Although the curve of quantum discord oscillates extremely frequently, the trend of its amplitude still has the same law to that of $\left\{0\succ_{cb}\right.$ and $\left\{1\succ_{cb}\right.$. Similarly, the trend is wavy, when $g_{\omega}$ is comparable to $g$, and becomes smoother, when the value of $g_{\omega}$ is getting smaller and smaller than $g$. And the trend is also highly synchronized with curves of $\left\{0\succ_{cb}\right.$ and $\left\{1\succ_{cb}\right.$: as the two curves of $\left\{0\succ_{cb}\right.$ and $\left\{1\succ_{cb}\right.$ gradually approach, the amplitude of the quantum discord becomes larger and larger; when the two curves intersect, the quantum discord reaches the maximum; then, the curves separate and gradually move away, and the quantum discord also decreases to $0$.
	
	Now we study the relationship between observed subsystem $\mathcal{A}$ (photon) and quantum discord. To this end, we define the following two states
	\begin{subequations}
		\label{eq:StatePhts}
		\begin{align}
			\left\{0\succ_{pht}\right.&=\sum_{\substack{p_1,p_2,m,l_1,l_2,\\L,k,p_1+p_2=0}}c''|p_1\rangle|p_2\rangle|m\rangle|l_1\rangle|l_2\rangle|L\rangle|k\rangle\label{eq:StatePht0}\\
			\left\{1\succ_{pht}\right.&=\sum_{\substack{p_1,p_2,m,l_1,l_2,\\L,k,p_1+p_2\geq1}}c'''|p_1\rangle|p_2\rangle|m\rangle|l_1\rangle|l_2\rangle|L\rangle|k\rangle\label{eq:StatePht1}
		\end{align}
	\end{subequations}
	where $c'',\ c'''$ are normalization factors, which depends on $p_1,\ p_2,\ m,\ l_1,\ l_2,\ L,\ k$. $\left\{0\succ_{pht}\right.$ --- there is no free photon in the system ($p_1+p_2=0$); $\left\{1\succ_{pht}\right.$ --- there is at least one photon in the system ($p_1+p_2\geq1$).
	
	Comparing (c) and (f) of Fig. \ref{fig:ClosedSystemWithTunneling}, it shows that $\left\{0\succ_{pht}\right.$ and $\left\{1\succ_{pht}\right.$ are synchronized with quantum discord. It is precisely because in the part of observed subsystem $\mathcal{A}$ appear at the same time the two situations: without photons and with photons, and these two situations interact with substance subsystem $\mathcal{B}$ to generate quantum correlation.
	
	\subsubsection{Without tunneling effect} 
	\label{subsubsec:WithoutTunneling}
	
	We assume that the quantum tunneling effect of the nuclei does not exist, that is, $\zeta=0$. In the absence of quantum tunneling effect, we consider two atoms that are initially close together so that they can interact to form a covalent bond. At this time, the initial states is also $|\Psi_{initial}\rangle$, but two atoms cannot move to different cavities without tunneling effect. 
	
	Compared with the results in Fig. \ref{fig:ClosedSystemWithTunneling}, the oscillations of quantum discord are no longer violent in Fig. \ref{fig:ClosedSystemWithoutTunneling}. And we find that the oscillations of $\left\{0\succ_{cb}\right.$, $\left\{1\succ_{cb}\right.$ and quantum discord become irregular. The curves of states $\left\{0\succ_{cb}\right.$ and $\left\{1\succ_{cb}\right.$ are no longer synchronized with the curve of quantum discord, but the curves of states $\left\{0\succ_{pht}\right.$ and $\left\{1\succ_{pht}\right.$ are still synchronized with it. This shows that states $\left\{0\succ_{pht}\right.$ and $\left\{1\succ_{pht}\right.$ can more accurately describe the relationship between quantum states and quantum discord.
	
	Now we fit all the curves about quantum discord in Fig. \ref{fig:ClosedSystemWithTunneling} and Fig. \ref{fig:ClosedSystemWithoutTunneling} to the $sin$-like function fitting process, and get Figs. \ref{fig:FittingFig4} and \ref{fig:FittingFig5}. Comparing Fig. \ref{fig:FittingFig5} with Fig. \ref{fig:FittingFig4}, we can obviously find when there is no tunneling effect, the amplitude of quantum discord becomes lower, and the period becomes shorter. It is very clear from Fig. \ref{fig:Comparison} that the period varies with $g_{\omega}$ in both cases: the larger the $g_{\omega}$, the shorter the period, and the period is always lower in the case of tunneling effect than that in the case without tunneling effect.
	
	\subsection{Open system} 
	\label{subsec:OpenSystem}
	
	\begin{figure}
		\begin{center}
		\includegraphics[width=0.48\textwidth]{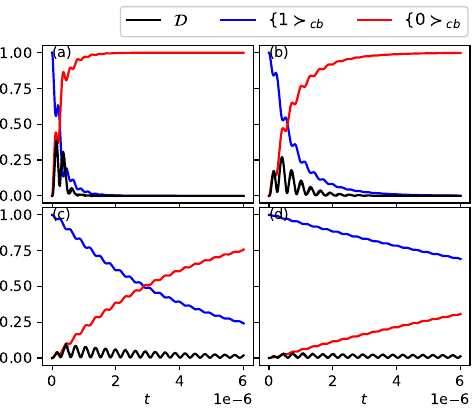}
		\end{center}
		\caption{(online color) {\it Impact of covalent bond strength on quantum discord dynamics with tunneling effect.} Here dissipative rate $\gamma=g$, thus dissipative dynamics is obtained. We assumed that the values of strengths $g_{\Omega^{\uparrow}}$ and $g_{\Omega^{\downarrow}}$ are equal to $g$, and tunneling effect is allowed, thus $\zeta=g$. We study the effect of different values of $g_{\omega}$ from large to small on the evolution and quantum discord dynamics: (a) $g_{\omega}=g$, (b) $g_{\omega}=0.5g$, (c) $g_{\omega}=0.2g$, (d) $g_{\omega}=0.1g$.}
		\label{fig:ImpactStrengthWithTunneling}
	\end{figure}
	
	\begin{figure}
		\begin{center}
		\includegraphics[width=0.48\textwidth]{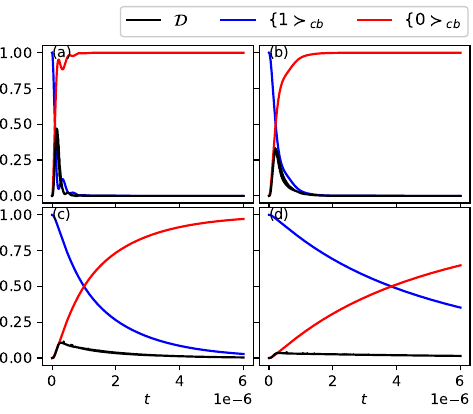}
		\end{center}
		\caption{(online color) {\it Impact of covalent bond strength on quantum discord dynamics without tunneling effect.} Dissipative dynamics is also obtained. However, the tunneling effect is forbidden, thus $\zeta=0$. In this case, we again study the effect of different values of $g_{\omega}$ on the evolution and quantum discord dynamics: (a) $g_{\omega}=g$, (b) $g_{\omega}=0.5g$, (c) $g_{\omega}=0.2g$, (d) $g_{\omega}=0.1g$.}
		\label{fig:ImpactStrengthWithoutTunneling}
	\end{figure}
	
	\begin{figure}
		\begin{center}
		\includegraphics[width=0.48\textwidth]{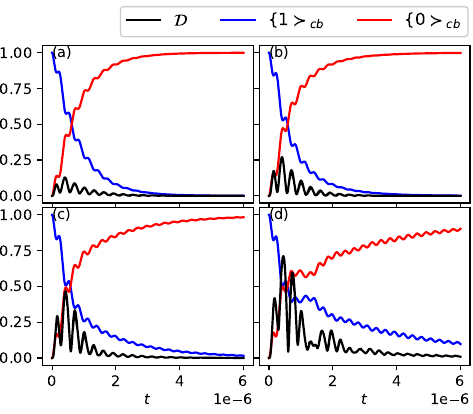}
		\end{center}
		\caption{(online color) {\it Impact of dissipation intensities on quantum discord dynamics with tunneling effect.} Dissipative dynamics is obtained. We assumed that the values of strengths $g_{\Omega^{\uparrow}}$, $g_{\Omega^{\downarrow}}$ and $\zeta$ are equal to $g$. And $g_{\omega}=0.5g$. We study the effect of different values of $\gamma$ from large to small on the evolution and quantum discord dynamics: (a) $\gamma=2g$, (b) $\gamma=g$, (c) $\gamma=0.5g$, (d) $\gamma=0.2g$.}
		\label{fig:ImpactIntensityWithTunneling}
	\end{figure}
	
	\begin{figure}
		\begin{center}
		\includegraphics[width=0.48\textwidth]{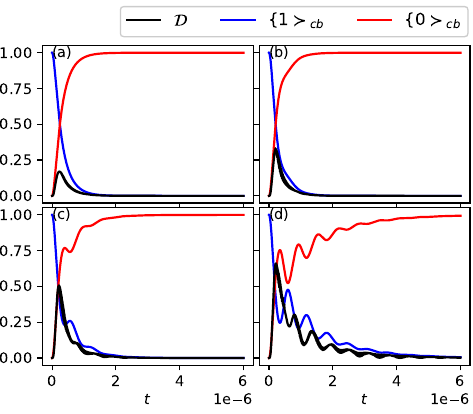}
		\end{center}
		\caption{(online color) {\it Impact of dissipation intensities on quantum discord dynamics without tunneling effect.} Dissipative dynamics is obtained. We assumed that the values of strengths $g_{\Omega^{\uparrow}}$, $g_{\Omega^{\downarrow}}$ are equal to $g$, and $g_{\omega}=0.5g$. However, $\zeta=0$. We again study the effect of different values of $\gamma$ from large to small on the evolution and quantum discord dynamics: (a) $\gamma=2g$, (b) $\gamma=g$, (c) $\gamma=0.5g$, (d) $\gamma=0.2g$.}
		\label{fig:ImpactIntensityWithoutTunneling}
	\end{figure}
	
	Now we focus on the dissipation dynamics in Markovian open system.
	
	\subsubsection{Impact of covalent bond strength on quantum discord dynamics}
	\label{subsubsec:ImpactStrength}
	
	We study the effect of different values of $g_{\omega}$ on quantum discord dynamics. We fix the value of all dissipation intensities: $\gamma_{\Omega^{\uparrow}}=\gamma_{\Omega^{\downarrow}}=\gamma_{\omega}=g$. At the same time, we also consider two cases: with tunneling effect and without tunneling effect. First, we consider the case of the tunneling effect. From Fig. \ref{fig:ImpactStrengthWithTunneling}, we can find that in an open system, the quantum discord will gradually weaken to $0$ as time increases. During this process, the curve for state $\left\{0\succ_{cb}\right.$ gradually rises and reaches $1$, which means that the quantum system reaches a stable molecular state and the quantum correlation disappears. And we found that the smaller $g_{\omega}$ is, the smaller the amplitude of the quantum discord is, but the decay speed of quantum discord is slower.
	
	Then, we focus on the case without tunneling effect. From Fig. \ref{fig:ImpactStrengthWithoutTunneling}, we can find the similar result as above. However, comparing Fig. \ref{fig:ImpactStrengthWithoutTunneling} with Fig. \ref{fig:ImpactStrengthWithTunneling}, when the quantum tunneling effect of atomic nuclei is not considered, the curve of quantum discord will be smoother than the case with tunneling effect.
	
	\subsubsection{Impact of dissipation intensities on quantum discord dynamics}
	\label{subsubsec:ImpactIntensity}
	
	Now we study the effect of different values of dissipation intensities $\gamma$ on quantum discord dynamics. We fix the value of covalent bond strength: $g_{\omega}=0.5g$. For case of tunneling effect, from Fig. \ref{fig:ImpactIntensityWithTunneling}, we can find that the quantum discord will also gradually weaken to $0$ as time increases. However, the difference from Fig. \ref{fig:ImpactIntensityWithTunneling} is that in Fig. \ref{fig:ImpactIntensityWithTunneling}, the smaller $\gamma$ is, the larger the amplitude of the quantum discord is. In other words, the coupling strength $g$ has a positive effect on the value of quantum discord, while the dissipation intensity $\gamma$ has a negative effect on it.
	
	For case without tunneling effect, from Fig. \ref{fig:ImpactIntensityWithoutTunneling}, we can find the similar result as in Fig. \ref{fig:ImpactIntensityWithTunneling}. However, comparing Fig. \ref{fig:ImpactIntensityWithoutTunneling} with Fig. \ref{fig:ImpactIntensityWithTunneling}, we can find that when the quantum tunneling effect of atomic nuclei is not considered, the curve of quantum discord will be smoother just like the case of Fig. \ref{fig:ImpactIntensityWithoutTunneling}. Therefore, we can conclude that when the tunneling effect disappears, the quantum discord will tend to be smooth in both cases.

	\section{Concluding discussion and future work} 
	\label{sec:ConcluFuture}
	
	In this paper, we have studied the quantum discord dynamics between light and matter via two-qubit measurement in a seven-qubit quantum system, and derived some analytical results of it
	
	\begin{itemize}
		\item In Sec. \ref{subsubsec:WithTunneling}, we have studied the quantum discord dynamics in closed system with tunneling effect. The trend of amplitude of quantum discord is highly synchronized with quantum system evolution. And it is wavy, when $g_{\omega}$ is comparable to $g$, and becomes smoother, when the value of $g_{\omega}$ is getting smaller and smaller than $g$. Photonic states $\left\{0\succ_{pht}\right.$ and $\left\{1\succ_{pht}\right.$ also synchronize with quantum discord.
		\item In Sec. \ref{subsubsec:WithoutTunneling}, we removed the tunneling effect. We have found that states $\left\{0\succ_{pht}\right.$ and $\left\{1\succ_{pht}\right.$ (still synchronized with quantum discord) can more accurately describe the relationship between unitary evolution and quantum discord than states $\left\{0\succ_{cb}\right.$ and $\left\{1\succ_{cb}\right.$ (no longer synchronized with quantum discord). When tunneling effect is absent, the amplitude of quantum discord becomes lower, and the period becomes shorter than that in the case of tunneling effect. 
		\item In Sec. \ref{subsubsec:ImpactIntensity}, consideration is given to the open system. We first focus on the effect of coupling strength on quantum discord. We have found that the coupling strength $g_{\omega}$ has negative effect on the existence of quantum discord: the larger $g_{\omega}$ is, the faster the decay speed of quantum discord is. And for the amplitude of quantum discord, $g_{\omega}$ has the positive effect on it, that is to say, the larger $g_{\omega}$ is, the higher the initial amplitude. Then, we removed the tunneling effect, the smoother curves are obtained.
		\item In Sec. \ref{subsubsec:ImpactIntensity}, we now focus on the effect of dissipation intensity on quantum discord. As same as the coupling strength, the dissipation intensity $\gamma$ has negative effect on the existence of quantum discord. However, for the amplitude, $\gamma$ has the negative effect on it, that is to say, the larger $\gamma$ is, the weaker the initial amplitude. This is because the larger the $\gamma$, the faster the photons and phonon escape, therefore, the smaller the quantum discord between photons and matter, and the faster the system will reach a stable hydrogen molecular structure. Then, we removed the tunneling effect, the smoother curves are also obtained just like the case of Sec. \ref{subsubsec:ImpactIntensity}.
	\end{itemize}
	
	Although the target model is rough, our goal is to try to analyze the quantum correlation (quantum discord) between light and matter in complex multi-qubit quantum systems, rather than simple two-qubit quantum systems. The first task is to dedicate to identifying the regularity of quantum correlation and study the factors that affect it as the basis for future research on more complex quantum systems. The ultimate goal is to realize the study of quantum discord of system of dozens and even hundreds of qubits in the future. These complicate systems include chemical or biological models of some macromolecules.
	
	\begin{acknowledgments}
	The reported study was funded by China Scholarship Council, project number 202108090483.
	\end{acknowledgments}

\appendix

	\section{Two-qubit von Neumann projective measurement}
	\label{appx:TwoQubitvonNeumann}

	The von Neumann measurement projective basis for a Hilbert space with dimension $4$ (two qubits) can be written as
	\begin{equation}
		\label{appxeq:MeasurementvonNeumann2Qubits}
		\begin{aligned}
			\{|b_k\rangle\}&=\{cos\theta|0\rangle+sin\theta|1\rangle,sin\theta|0\rangle-cos\theta|1\rangle\}\\
			&\otimes\{cos\theta'|0\rangle+sin\theta'|1\rangle,sin\theta'|0\rangle-cos\theta'|1\rangle\}
		\end{aligned}
	\end{equation}
where $k=0,\ 1,\ 2,\ 3$ and $\theta,\ \theta'\in[0,\frac{\pi}{2}]$. $\theta$ corresponds to the first qubit and $\theta'$ corresponds to the second qubit. Then, eliminating the vector product in the Eq. \ref{appxeq:MeasurementvonNeumann2Qubits}, we can get the four following expressions
	\begin{subequations}
		\label{appxeq:BasesvonNeumann2Qubits}
		\begin{align}
			|b_0\rangle&=cos\theta cos\theta'|00\rangle+cos\theta sin\theta'|01\rangle\nonumber\\
			&+sin\theta cos\theta'|10\rangle+sin\theta sin\theta'|11\rangle\label{appxeq:Base0}\\
			|b_1\rangle&=sin\theta cos\theta'|00\rangle+sin\theta sin\theta'|01\rangle\nonumber\\
			&-cos\theta cos\theta'|10\rangle-cos\theta sin\theta'|11\rangle\label{appxeq:Base1}\\
			|b_2\rangle&=cos\theta sin\theta'|00\rangle-cos\theta cos\theta'|01\rangle\nonumber\\
			&+sin\theta sin\theta'|10\rangle-sin\theta cos\theta'|11\rangle\label{appxeq:Base2}\\
			|b_3\rangle&=sin\theta sin\theta'|00\rangle-sin\theta cos\theta'|01\rangle\nonumber\\
			&-cos\theta sin\theta'|10\rangle+cos\theta cos\theta'|11\rangle\label{appxeq:Base3}
		\end{align}
	\end{subequations}
Now the projection operators for full two-qubit Hilbert space are defined as follows
	\begin{widetext}
		\begin{subequations}
			\label{appxeq:Projecteurs2Qubits}
			\begin{align}
				\Pi_0^{\mathcal{A}}&=|b_0\rangle\langle b_0|\nonumber\\
				&=cos^2\theta cos^2\theta'|00\rangle\langle00|+cos^2\theta cos\theta'sin\theta'|00\rangle\langle01|+cos\theta cos^2\theta'sin\theta|00\rangle\langle10|+cos\theta cos\theta'sin\theta sin\theta'|00\rangle\langle11|\nonumber\\
				&+cos^2\theta sin\theta'cos\theta'|01\rangle\langle00|+cos^2\theta sin^2\theta'|01\rangle\langle01|+cos\theta sin\theta'sin\theta cos\theta'|01\rangle\langle10|+cos\theta sin^2\theta'sin\theta|01\rangle\langle11|\nonumber\\
				&+sin\theta cos^2\theta'cos\theta|10\rangle\langle00|+sin\theta cos\theta'cos\theta sin\theta'|10\rangle\langle01|+sin^2\theta cos^2\theta'|10\rangle\langle10|+sin^2\theta cos\theta'sin\theta'|10\rangle\langle11|\nonumber\\
				&+sin\theta sin\theta'cos\theta cos\theta'|11\rangle\langle00|+sin\theta sin^2\theta'cos\theta|11\rangle\langle01|+sin^2\theta sin\theta'cos\theta'|11\rangle\langle10|+sin^2\theta sin^2\theta'|11\rangle\langle11|\label{appxeq:Projecteur2Qubits0}\\
				\Pi_1^{\mathcal{A}}&=|b_1\rangle\langle b_1|\nonumber\\
				&=sin^2\theta cos^2\theta'|00\rangle\langle00|+sin^2\theta cos\theta'sin\theta'|00\rangle\langle01|-sin\theta cos^2\theta'cos\theta|00\rangle\langle10|-sin\theta cos\theta'cos\theta sin\theta'|00\rangle\langle11|\nonumber\\
				&+sin^2\theta sin\theta'cos\theta'|01\rangle\langle00|+sin^2\theta sin^2\theta'|01\rangle\langle01|-sin\theta sin\theta'cos\theta cos\theta'|01\rangle\langle10|-sin\theta sin^2\theta'cos\theta|01\rangle\langle11|\nonumber\\
				&-cos\theta cos^2\theta'sin\theta|10\rangle\langle00|-cos\theta cos\theta'sin\theta sin\theta'|10\rangle\langle01|+cos^2\theta cos^2\theta'|10\rangle\langle10|+cos^2\theta cos\theta'sin\theta'|10\rangle\langle11|\nonumber\\
				&-cos\theta sin\theta'sin\theta cos\theta'|11\rangle\langle00|-cos\theta sin^2\theta'sin\theta|11\rangle\langle01|+cos^2\theta sin\theta'cos\theta'|11\rangle\langle10|+cos^2\theta sin^2\theta'|11\rangle\langle11|\label{appxeq:Projecteur2Qubits1}\\
				\Pi_2^{\mathcal{A}}&=|b_2\rangle\langle b_2|\nonumber\\
				&=cos^2\theta sin^2\theta'|00\rangle\langle00|-cos^2\theta sin\theta'cos\theta'|00\rangle\langle01|+cos\theta sin^2\theta'sin\theta|00\rangle\langle10|-cos\theta sin\theta'sin\theta cos\theta'|00\rangle\langle11|\nonumber\\
				&-cos^2\theta cos\theta'sin\theta'|01\rangle\langle00|+cos^2\theta cos^2\theta'|01\rangle\langle01|-cos\theta cos\theta'sin\theta sin\theta'|01\rangle\langle10|+cos\theta cos^2\theta'sin\theta|01\rangle\langle11|\nonumber\\
				&+sin\theta sin^2\theta'cos\theta|10\rangle\langle00|-sin\theta sin\theta'cos\theta cos\theta'|10\rangle\langle01|+sin^2\theta sin^2\theta'|10\rangle\langle10|-sin^2\theta sin\theta'cos\theta'|10\rangle\langle11|\nonumber\\
				&-sin\theta cos\theta'cos\theta sin\theta'|11\rangle\langle00|+sin\theta cos^2\theta'cos\theta|11\rangle\langle01|-sin^2\theta cos\theta'sin\theta'|11\rangle\langle10|+sin^2\theta cos^2\theta'|11\rangle\langle11|\label{appxeq:Projecteur2Qubits2}\\
				\Pi_3^{\mathcal{A}}&=|b_3\rangle\langle b_3|\nonumber\\
				&=sin^2\theta sin^2\theta'|00\rangle\langle00|-sin^2\theta sin\theta'cos\theta'|00\rangle\langle01|-sin\theta sin^2\theta'cos\theta|00\rangle\langle10|+sin\theta sin\theta'cos\theta cos\theta'|00\rangle\langle11|\nonumber\\
				&-sin^2\theta cos\theta'sin\theta'|01\rangle\langle00|+sin^2\theta cos^2\theta'|01\rangle\langle01|+sin\theta cos\theta'cos\theta sin\theta'|01\rangle\langle10|-sin\theta cos^2\theta'cos\theta|01\rangle\langle11|\nonumber\\
				&-cos\theta sin^2\theta'sin\theta|10\rangle\langle00|+cos\theta sin\theta'sin\theta cos\theta'|10\rangle\langle01|+cos^2\theta sin^2\theta'|10\rangle\langle10|-cos^2\theta sin\theta'cos\theta'|10\rangle\langle11|\nonumber\\
				&+cos\theta cos\theta'sin\theta sin\theta'|11\rangle\langle00|-cos\theta cos^2\theta'sin\theta|11\rangle\langle01|-cos^2\theta cos\theta'sin\theta'|11\rangle\langle10|+cos^2\theta cos^2\theta'|11\rangle\langle11|\label{appxeq:Projecteur2Qubits3}
			\end{align}
		\end{subequations}
	\end{widetext}
In order to reduce modeling difficulty and time cost, we can assume $\theta=\theta'$. Thus,
	\begin{widetext}
	 	\begin{subequations}
			\label{appxeq:ProjSimples2Qubits}
			\begin{align}
				\Pi_0^{\mathcal{A}}&=C_0|00\rangle\langle00|+C_1\left(|00\rangle\langle01|+|01\rangle\langle00|+|00\rangle\langle10|+|10\rangle\langle00|\right)+C_3\left(|01\rangle\langle11|+|11\rangle\langle01|+|10\rangle\langle11|+|11\rangle\langle10|\right)\nonumber\\
				&+C_2\left(|01\rangle\langle01|+|10\rangle\langle10|+|00\rangle\langle11|+|11\rangle\langle00|+|01\rangle\langle10|+|10\rangle\langle01|\right)+C_4|11\rangle\langle11|\label{appxeq:ProjSimple2Qubits0}\\
				\Pi_1^{\mathcal{A}}&=C_0|10\rangle\langle10|+C_1\left(-|00\rangle\langle10|-|10\rangle\langle00|+|10\rangle\langle11|+|11\rangle\langle10|\right)+C_3\left(|00\rangle\langle01|+|01\rangle\langle00|-|01\rangle\langle11|-|11\rangle\langle01|\right)\nonumber\\
				&+C_2\left(|00\rangle\langle00|+|11\rangle\langle11|-|00\rangle\langle11|-|11\rangle\langle00|-|01\rangle\langle10|-|10\rangle\langle01|\right)+C_4|01\rangle\langle01|\label{appxeq:ProjSimple2Qubits1}\\
				\Pi_2^{\mathcal{A}}&=C_0|01\rangle\langle01|+C_1\left(-|00\rangle\langle01|-|01\rangle\langle00|+|01\rangle\langle11|+|11\rangle\langle01|\right)+C_3\left(|00\rangle\langle10|+|10\rangle\langle00|-|10\rangle\langle11|-|11\rangle\langle10|\right)\nonumber\\
				&+C_2\left(|00\rangle\langle00|+|11\rangle\langle11|-|00\rangle\langle11|-|11\rangle\langle00|-|01\rangle\langle10|-|10\rangle\langle01|\right)+C_4|10\rangle\langle10|\label{appxeq:ProjSimple2Qubits2}\\
				\Pi_3^{\mathcal{A}}&=C_0|11\rangle\langle11|+C_1\left(-|01\rangle\langle11|-|11\rangle\langle01|-|10\rangle\langle11|-|11\rangle\langle10|\right)+C_3\left(-|00\rangle\langle01|-|01\rangle\langle00|-|00\rangle\langle10|-|10\rangle\langle00|\right)\nonumber\\
				&+C_2\left(|01\rangle\langle01|+|10\rangle\langle10|+|00\rangle\langle11|+|11\rangle\langle00|+|01\rangle\langle10|+|10\rangle\langle01|\right)+C_4|00\rangle\langle00|\label{appxeq:ProjSimple2Qubits3}
			\end{align}
		\end{subequations}
	\end{widetext}
	where
	\begin{equation}
		\label{appxeq:ConstantCs}
		\begin{aligned}
			C_0&=cos^4\theta\\
			C_1&=cos^3\theta sin\theta\\
			C_2&=cos^2\theta sin^2\theta\\
			C_3&=cos^1\theta sin^3\theta\\
			C_4&=sin^4\theta
		\end{aligned}
	\end{equation}
	
	In addition to the above method, we can construct these projection operators $\Pi^{\mathcal{A}}_k$ in other ways. For example, we can first construct single-qubit projection operators for different qubits, and then combine these projection operators through the tensor product. In this paper, we have two single-qubit projection operators $Pr_0,\ Pr_1$ corresponds to the first qubit and another two single-qubit projection operators $Pr_0',\ Pr_1'$ corresponds to the second qubit, which are defined as follows
	\begin{subequations}
		\label{appxeq:Projecteurs1Qubit}
		\begin{align}
			Pr_0&=cos^2\theta|0\rangle\langle0|+cos\theta sin\theta|0\rangle\langle1|\nonumber\\
			&+sin\theta cos\theta|1\rangle\langle0|+sin^2\theta|1\rangle\langle1|\label{appxeq:Projecteur1Qubit0}\\
			Pr_1&=sin^2\theta|0\rangle\langle0|-sin\theta cos\theta|0\rangle\langle1|\nonumber\\
			&-cos\theta sin\theta|1\rangle\langle0|+cos^2\theta|1\rangle\langle1|\label{appxeq:Projecteur1Qubit1}\\
			Pr_0'&=cos^2\theta'|0\rangle\langle0|+cos\theta' sin\theta'|0\rangle\langle1|\nonumber\\
			&+sin\theta' cos\theta'|1\rangle\langle0|+sin^2\theta'|1\rangle\langle1|\label{appxeq:Projecteur1Qubit0'}\\
			Pr_1'&=sin^2\theta'|0\rangle\langle0|-sin\theta' cos\theta'|0\rangle\langle1|\nonumber\\
			&-cos\theta' sin\theta'|1\rangle\langle0|+cos^2\theta'|1\rangle\langle1|\label{appxeq:Projecteur1Qubit1'}
		\end{align}
	\end{subequations}
Now the projection operators for full two-qubit space are simply defined as
	\begin{subequations}
		\label{appxeq:ProjNew2Qubits}
		\begin{align}
			\Pi_0^{\mathcal{A}}&=Pr_0\otimes Pr_0'\\
			\Pi_1^{\mathcal{A}}&=Pr_1\otimes Pr_0'\\
			\Pi_2^{\mathcal{A}}&=Pr_0\otimes Pr_1'\\
			\Pi_3^{\mathcal{A}}&=Pr_1\otimes Pr_1'
		\end{align}
	\end{subequations}
	
	Besides, the von Neumann projective measurement basis in Eq. \eqref{appxeq:MeasurementvonNeumann2Qubits} can be extended to the general form
	\begin{equation}
		\begin{aligned}
			\label{eq:MeasurementGeneral2Qubits}
			\{|b_k\rangle\}&=\{cos\theta|0\rangle+sin\theta e^{i\varphi}|1\rangle,sin\theta e^{i\varphi}|0\rangle-cos\theta|1\rangle\}\\
			&\otimes\{cos\theta'|0\rangle+sin\theta'e^{i\varphi'}|1\rangle,sin\theta'e^{i\varphi'}|0\rangle-cos\theta'|1\rangle\}
		\end{aligned}
	\end{equation}
	where $\varphi,\ \varphi'\in[0,2\pi]$. Similarly, in order to reduce modeling difficulty and time cost, we can also assume $\varphi=\varphi'$.
	
	\section{Second quantization}
	\label{appx:SecondQuantization}

	We introduce the second quantization, also known as the occupation number representation \cite{Dirac1927, Fock1932}, to prevent the difficulty that antisymmetrization causes from becoming more complicated. In this approach, the quantum many-body states are represented in the Fock state basis, which are constructed by filling up each single-particle state with a certain number of identical particles
	\begin{equation}
		\label{appxeq:FockState}
		|Fock\rangle=|n_1,n_2,n_3,\cdots,n_{\alpha},\cdots\rangle
	\end{equation}
	
	In the single-particle state $|\alpha\rangle$, it signifies that there are $n_{\alpha}$ particles. The total number of particles $N$ is equal to the sum of the occupation numbers, or $\sum_{\alpha}n_{\alpha}=N$. Due to the Pauli exclusion principle, the occupancy number $n_{\alpha}$ for fermions can only be $0$ or $1$ but it can be any non-negative integer for bosons. The many-body Hilbert space, also known as Fock space, is completely based on all of the Fock states. A linear collection of Fock states can be used to express any generic quantum many-body state. The creation and annihilation operators are introduced in the second quantization formalism to construct and handle the Fock states, giving researchers studying the quantum many-body theory useful tools.
	
	\section{Operators}
	\label{appx:Operators}
	
	On a $p$-photons state (or $m$-phonons state), the photon annihilation and creation operators $a$ and $a^{\dag}$ are described as
	\begin{equation}
		\label{appxeq:PhotonOperators}
		\begin{aligned}
			&if\ p>0,\ \left\{
				\begin{aligned}
				&a|p\rangle=\sqrt{p}|p-1\rangle,\\
				&a^{\dag}|p\rangle=\sqrt{p+1}|p+1\rangle,
				\end{aligned}
				\right
				.\\
			&if\ p=0,\ \left \{
				\begin{aligned}
					&a|0\rangle=0,\\
					&a^{\dag}|0\rangle=|1\rangle.
				\end{aligned}
				\right
				.\\
		\end{aligned}
	\end{equation}
	Operators $a_{\Omega^{\uparrow}}$, $a_{\Omega^{\downarrow}}$, $a_{\omega}$ and their hermitian conjugate operators all obey the rules in Eq. \eqref{appxeq:PhotonOperators}.
	
	The interaction of molecule with the electromagnetic field of the cavity, emitting or absorbing photon with mode $\Omega^{\uparrow,\downarrow}$, is described as
	\begin{equation}
		\label{appxeq:InteractionMolecule}
		\begin{aligned}
			&\sigma_{\Omega^{\uparrow}}|1\rangle_{\Phi_1}^{\uparrow}|0\rangle_{\Phi_0}^{\uparrow}=|0\rangle_{\Phi_1}^{\uparrow}|1\rangle_{\Phi_0}^{\uparrow},\\
			&\sigma_{\Omega^{\uparrow}}^{\dag}|0\rangle_{\Phi_1}^{\uparrow}|1\rangle_{\Phi_0}^{\uparrow}=|1\rangle_{\Phi_1}^{\uparrow}|0\rangle_{\Phi_0}^{\uparrow},\\
			&\sigma_{\Omega^{\downarrow}}|1\rangle_{\Phi_1}^{\downarrow}|0\rangle_{\Phi_0}^{\downarrow}=|0\rangle_{\Phi_1}^{\downarrow}|1\rangle_{\Phi_0}^{\downarrow},\\
			&\sigma_{\Omega^{\downarrow}}^{\dag}|0\rangle_{\Phi_1}^{\downarrow}|1\rangle_{\Phi_0}^{\downarrow}=|1\rangle_{\Phi_1}^{\downarrow}|0\rangle_{\Phi_0}^{\downarrow}.
		\end{aligned}
	\end{equation}
	
	The covalent bond's formation and break operators have following form
	\begin{equation}
		\label{appxeq:BondOperators}
		\begin{aligned}
			&\sigma_{\omega}|1\rangle_{cb}=|0\rangle_{cb},\\
			&\sigma_{\omega}^{\dag}|0\rangle_{cb}=|1\rangle_{cb}.
		\end{aligned}
	\end{equation}
	
	The nuclei's tunnelling operators have following form
	\begin{equation}
		\label{appxeq:TunnellingOperators}
		\begin{aligned}
			&\sigma_n|1\rangle_n=|0\rangle_n,\\
			&\sigma_n^{\dag}|0\rangle_n=|1\rangle_n.
		\end{aligned}
	\end{equation}
	
	The matrix form of annihilation and creation operators in this paper is described as follows
	\begin{subequations}
		\label{appxeq:OperatorsA}
		\begin{align}
			a&=\begin{array}{c@{\hspace{-5pt}}l}
			 \begin{array}{c}
			 	|0\rangle \\
			 	|1\rangle \\
			 	|2\rangle \\
			 	\vdots \\
			 	\vdots \\
			 	|p-2\rangle \\
			 	|p-1\rangle \\
			 	|p\rangle \\
			 \end{array}
			 & \left(
			 \begin{array}{cccccccc}
			 	0 & 1 & 0 & \cdots & \cdots & 0 & 0 & 0 \\
			 	0 & 0 & \sqrt{2} & \cdots & \cdots & 0 & 0 & 0 \\
			 	0 & 0 & 0 & \cdots & \cdots & 0 & 0 & 0 \\
			 	\vdots & \vdots & \vdots & \ddots & \ddots & \vdots & \vdots & \vdots \\
			 	\vdots & \vdots & \vdots & \ddots & \ddots & \vdots & \vdots & \vdots \\
			 	0 & 0 & 0 & \cdots & \cdots & 0 & \sqrt{p-1} & 0 \\
			 	0 & 0 & 0 & \cdots & \cdots & 0 & 0 & \sqrt{p}\\
			 	0 & 0 & 0 & \cdots & \cdots & 0 & 0 & 0 \\
			 \end{array}
			 \right)
			\end{array}\label{appxeq:Annihilation}\\
			a^{\dag}&=\begin{array}{c@{\hspace{-5pt}}l}
			 \begin{array}{c}
			 	|0\rangle \\
			 	|1\rangle \\
			 	|2\rangle \\
			 	\vdots \\
			 	\vdots \\
			 	|p-2\rangle \\
			 	|p-1\rangle \\
			 	|p\rangle \\
			 \end{array}
			 & \left(
			 \begin{array}{cccccccc}
			 	0 & 0 & 0 & \cdots & \cdots & 0 & 0 & 0 \\
			 	1 & 0 & 0 & \cdots & \cdots & 0 & 0 & 0 \\
			 	0 & \sqrt{2} & 0 & \cdots & \cdots & 0 & 0 & 0 \\
			 	\vdots & \vdots & \vdots & \ddots & \ddots & \vdots & \vdots & \vdots \\
			 	\vdots & \vdots & \vdots & \ddots & \ddots & \vdots & \vdots & \vdots \\
			 	0 & 0 & 0 & \cdots & \cdots & 0 & 0 & 0 \\
			 	0 & 0 & 0 & \cdots & \cdots & \sqrt{p-1} & 0 & 0 \\
			 	0 & 0 & 0 & \cdots & \cdots & 0 & \sqrt{p} & 0 \\
			 \end{array}
			 \right)
			\end{array}\label{appxeq:Creation}
		\end{align}
	\end{subequations}
	And the matrix form of relaxation and excitation operators is described as follows
	\begin{subequations}
		\label{appxeq:OperatorsSigma}
		\begin{align}
			\sigma&=\begin{array}{c@{\hspace{-5pt}}l}
			 \begin{array}{c}
			 	|0\rangle \\
			 	|1\rangle \\
			 \end{array}
			 & \left(
			 \begin{array}{cc}
			 	0 & 1 \\
			 	0 & 0 \\
			 \end{array}
			 \right)
			\end{array}\label{appxeq:Relaxation}\\
			\sigma^{\dag}&=\begin{array}{c@{\hspace{-5pt}}l}
			 \begin{array}{c}
			 	|0\rangle \\
			 	|1\rangle \\
			 \end{array}
			 & \left(
			 \begin{array}{cc}
			 	0 & 0 \\
			 	1 & 0 \\
			 \end{array}
			 \right)
			\end{array}\label{appxeq:Excitation}
		\end{align}
	\end{subequations}

	\section{Quantum master equation}
	\label{appx:QME}

	The dynamics of system is described by solving the QME for the density matrix with the Lindblad operators of photon leakage from the cavity to external environment. The QME in the Markovian approximation for the density operator $\rho$ of the system takes the following form
	\begin{equation}
		\label{appxeq:QME}
		i\hbar\dot{\rho}=\mathcal{L}\left(\rho\right)=\left[H_{sys},\rho\right]+iL\left(\rho\right)
	\end{equation}
	where $\mathcal{L}\left(\rho\right)$ is Lindblad superoperator and $\left[H_{sys},\rho\right]=H_{sys}\rho-\rho H_{sys}$ is the commutator. We have a graph $\mathcal{K}$ of the potential photon dissipations between the states that are permitted. The edges and vertices of $\mathcal{K}$ represent the permitted dissipations and the states, respectively. Similar to this, $\mathcal{K}'$ is a graph of potential photon influxes that are permitted. $L\left(\rho\right)$ is as follows
	\begin{equation}
		\label{appxeq:LindbladOperator}
		L\left(\rho\right)=\sum_{k\in \mathcal{K}} L_k\left(\rho\right)+\sum_{k'\in \mathcal{K}'} L_{k'}\left(\rho\right)
	\end{equation}
	where $L_k\left(p\right)$ is the standard dissipation superoperator corresponding to the jump operator $A_k$ and taking as an argument on the density matrix $\rho$
	\begin{equation}
		\label{appxeq:DissSuper}
		L_k\left(\rho\right)=\gamma_k\left(A_k\rho A_k^{\dag}-\frac{1}{2}\left\{\rho, A_k^{\dag}A_k\right\}\right)
	\end{equation}
	where $\left\{\rho, A_k^{\dag}A_k\right\}=\rho A_k^{\dag}A_k + A_k^{\dag}A_k\rho$ is the anticommutator. 
	The term $\gamma_k$ refers to the overall spontaneous emission rate for photons for $k\in \mathcal{K}$ caused by photon leakage from the cavity to the external environment. Similarly, $L_{k'}\left(p\right)$ is the standard influx superoperator, having the following form
	\begin{equation}
		\label{appxeq:InfluxSuper}
		L_{k'}\left(\rho\right)=\gamma_{k'}\left(A_k^{\dag}\rho A_k-\frac{1}{2}\left\{\rho, A_kA_k^{\dag}\right\}\right)
	\end{equation}
	The total spontaneous influx rate for photon for $k'\in \mathcal{K}'$ is denoted by $\gamma_{k'}$.
	
	\section{Numerical method} 
	\label{appx:Method}
	
	\begin{figure}
		\centering
		\includegraphics[width=0.45\textwidth]{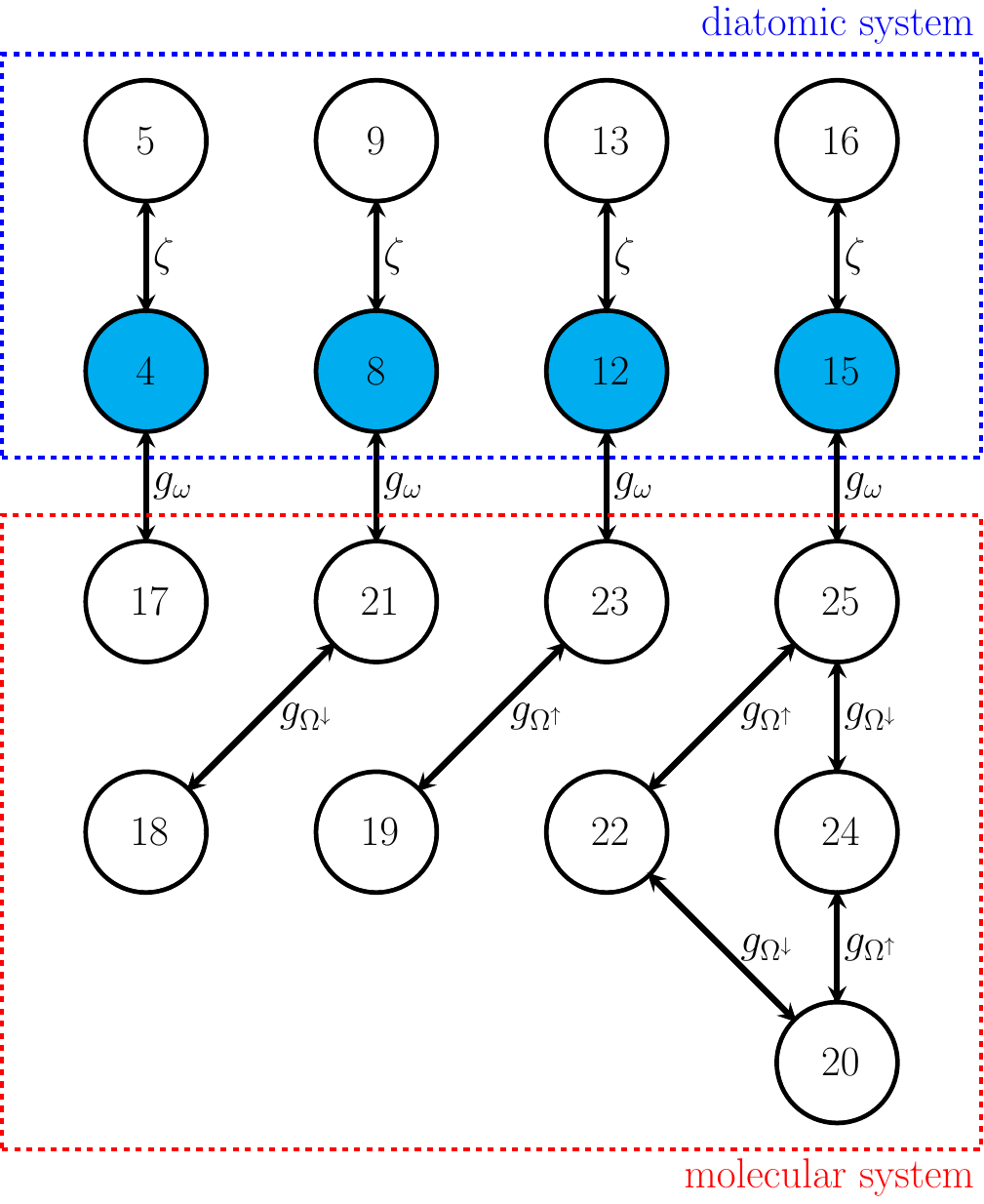}
		\caption{(online color) {\it Schematic diagram of evolution in the closed system.} Each circle represents a quantum state, and the number in the circle represents the index of the quantum state, which we can find in the Tab. \ref{Tab:States}. The four blue circles $\textcircled{4},\ \textcircled{8},\ \textcircled{12},\ \textcircled{15}$ represent the quantum states $|\Phi_0^{\uparrow}\Phi_0^{\downarrow}\rangle$, $|\Phi_0^{\uparrow}\Phi_1^{\downarrow}\rangle$, $|\Phi_1^{\uparrow}\Phi_0^{\downarrow}\rangle$ and $|\Phi_1^{\uparrow}\Phi_1^{\downarrow}\rangle$, which constitute the initial state. Black double-headed arrows represent interactions that connect two quantum states. Interactions include $g_{\Omega^{\uparrow}},\ g_{\Omega^{\downarrow}},\ g_{\omega},\ \zeta$. Transitions of system state can be achieved through these interactions. In a closed system, the state of the system moves from the initial state to other quantum states through various interactions, tunneling effect, etc. The circles within the blue dotted rectangle belong to diatomic system. The circles within the red dotted rectangle belong to molecular system.}
		\label{appxfig:EvolutionClosed}
	\end{figure}
	
	\begin{figure}
		\centering
		\includegraphics[width=0.45\textwidth]{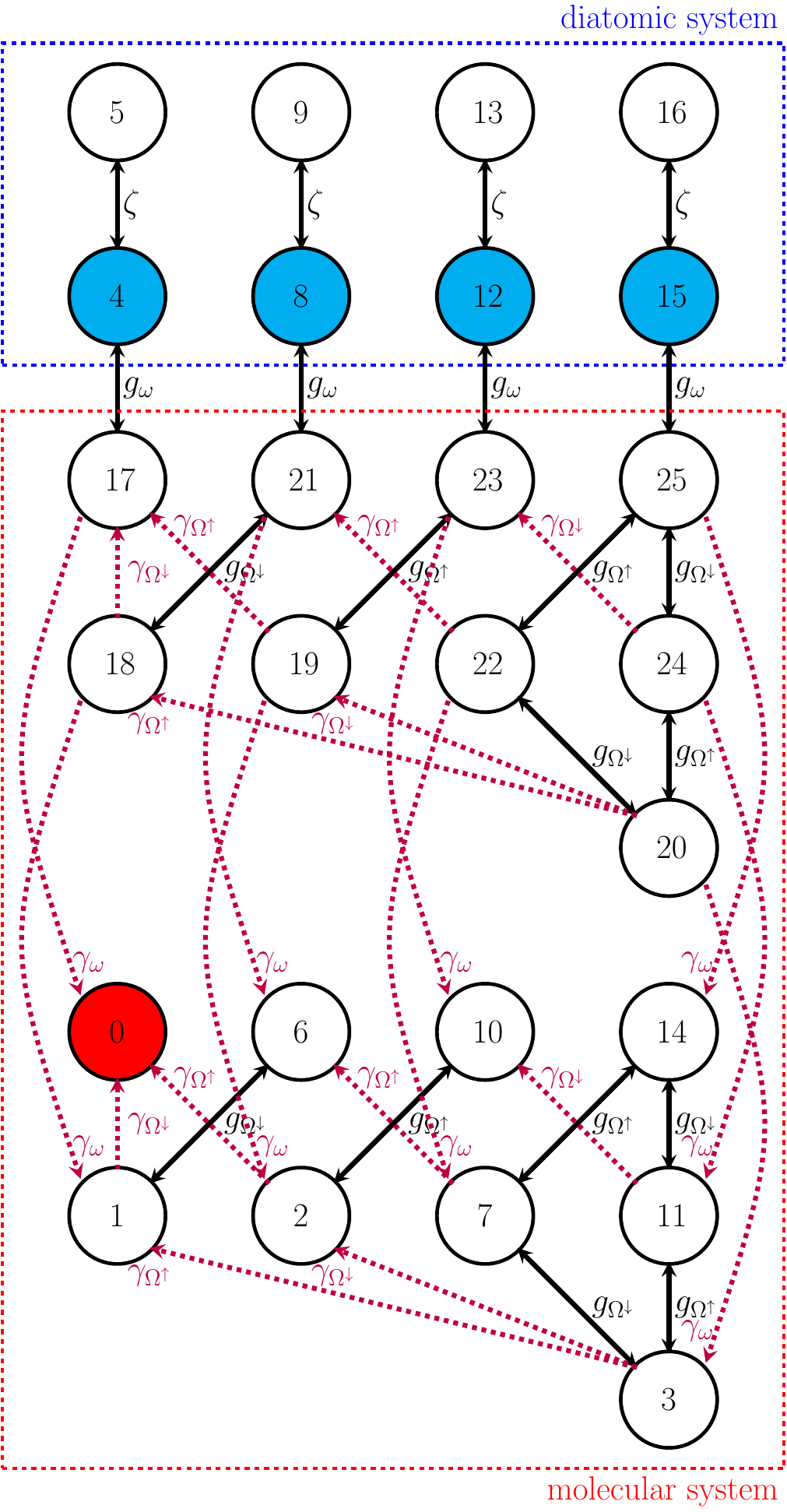}
		\caption{(online color) {\it Schematic diagram of evolution in the open system.} Compared with the closed system, the open system involve more quantum states due to dissipation effects. Obviously, open system is more complex than closed system. The purple dotted single-directional arrows represent the dissipation effects ($\gamma_{\Omega^{\uparrow}},\ \gamma_{\Omega^{\downarrow}},\ \gamma_{\omega}$). Also because of the escape of photons and phonons (we do not consider their influx), the energy of the entire system will eventually flow from the initial state to the final state $\textcircled{0}$, denoted by red circle. At this time, we obtain stable neutral hydrogen molecule.}
		\label{appxfig:EvolutionOpen}
	\end{figure}
	
	The solution $\rho\left(t\right)$ in Eq. \eqref{eq:QME} may be approximately found as a sequence of two steps: in the first step we make one step in the solution of the unitary part of Eq. \eqref{eq:QME}
	\begin{equation}
		\label{appxeq:UnitaryPart}
		\tilde{\rho}\left(t+dt\right)=exp\left({-\frac{i}{\hbar}H_{sys}dt}\right)\rho\left(t\right)exp\left(\frac{i}{\hbar}H_{sys}dt\right)
	\end{equation}
and in the second step, make one step in the solution of Eq. \eqref{eq:QME} with the commutator removed:
	\begin{equation}
		\label{appxeq:Solution}
		\rho\left(t+dt\right)=\tilde{\rho}\left(t+dt\right)+\frac{1}{\hbar}L\left(\tilde{\rho}(t+dt)\right)dt
	\end{equation}
	
	The main problem of quantum many-body physics is the fact that the Hilbert space grows exponentially with size, which we call the curse of dimensionality. In order to solve this problem, several schemes including the density matrix renormalization group (DMRG) method \cite{White1992, White1993} have been proposed. Our task is to describe a qualitative scenario of chemical dynamics, so we take the following method.
	
	We have the conventional technique known as tensor product for establishing Hamiltonian in Eq. \eqref{appxeq:UnitaryPart}. Through the use of the tensor product, we can directly establish the Hamiltonian with Eq. \eqref{eq:HamilBondPhonon}; however, the dimension of the Hamiltonian that results from this method is frequently very large and contains a lot of excess states that are not involved in evolution, particularly when the degree of freedom of the system is high. In this appendix, we will introduce the generator algorithm, which is based on the occupation number representation in Eq. \eqref{eq:SpaceBondPhonon}, and includes the following two steps
	\begin{itemize}
		\item generating and numbering potential evolution states involved in the evolution in accordance with the initial state and any its potential dissipative states that may be relevant in solving QME;
		\item establishing Hamiltonian with these states and potential interactions and dissipations among them.
	\end{itemize}
	Using this technique, we now eliminate the extra unnecessary states and obtain anew $\mathcal{C}'$ and $H_{sys}'$, where $\mathcal{C}'\subset\mathcal{C}$ and $dim\left(H_{sys}'\right)\leq dim\left(H_{sys}\right)$. In this paper, the $dim\left(H_{sys}'\right)=26$ is smaller than the $dim\left(H_{sys}\right)=2^{7}=128$. Complexity is reduced. The effectiveness of this reduction strategy increases with the increase of degree of freedom for multi-particle systems. Especially in complex macromolecular systems, the effect of reducing computational complexity is extremely obvious. More details about generator algorithm can be seen in \cite{Miao2023}.
	
	Schematic diagram of evolution in the closed system is shown in Fig. \ref{appxfig:EvolutionClosed}, and schematic diagram of evolution in the open system is shown in Fig. \ref{appxfig:EvolutionOpen}. All states in these two diagrams can be found in Tab. \ref{Tab:States}. We can clearly see that in a closed system, energy continuously flows between the diatomic system and the molecular system, while in an open system, the energy is ultimately stored in the molecular system.
	
	\section{Abbreviations and notations}
	\label{appx:AbbreviationsNotations}
	
	See Tab. \ref{appxtab:AbbreviationsNotations}.
	
	\onecolumngrid
		
	\begin{longtable}{ll}
	
		\caption{List of abbreviations and notations used in this paper.}\vspace{0.18cm}\\
		\label{appxtab:AbbreviationsNotations}\\
		\hline
		\multicolumn{1}{l}{\textbf{Abbreviations/Notations}} & \multicolumn{1}{l}{\textbf{Descriptions}}\vspace{0.18cm}\\
		\hline
		\endfirsthead
		
		\multicolumn{2}{c}{{\bfseries -- continued from previous page}}\vspace{0.18cm}\\
		\hline
		\multicolumn{1}{l}{\textbf{Abbreviations/Notations}} & \multicolumn{1}{l}{\textbf{Descriptions}}\vspace{0.18cm}\\
		\hline
		\endhead
		
		\hline
		\multicolumn{1}{l}{{Continued on next page}}\vspace{0.18cm}
		\endfoot
		
		\hline
		\endlastfoot
		
		QIP & Quantum information processing\vspace{0.18cm}\\
		QED & Quantum electrodynamics\vspace{0.18cm}\\
		USC & Ultrastrong-coupling\vspace{0.18cm}\\
		DSC & Deep strong coupling\vspace{0.18cm}\\
		SC & Strong coupling\vspace{0.18cm}\\
		JCM & Jaynes--Cummings model\vspace{0.18cm}\\
		TCM & Tavis--Cummings model\vspace{0.18cm}\\
		JCHM & Jaynes--Cummings--Hubbard model\vspace{0.18cm}\\
		TCHM & Tavis--Cummings--Hubbard model\vspace{0.18cm}\\
		QME & Quantum master equation\vspace{0.18cm}\\
		RWA & Rotating wave approximation\vspace{0.18cm}\\
		DMRG & Density matrix renormalization group\vspace{0.18cm}\\
		AO & Atomic orbital\vspace{0.18cm}\\
		MO & Molecular orbital\vspace{0.18cm}\\
		$A$ & Variable\vspace{0.18cm}\\
		$B$ & Variable\vspace{0.18cm}\\
		$I$ & Mutual information\vspace{0.18cm}\\
		$J$ & Mutual information\vspace{0.18cm}\\
		$H$ & \tabincell{l}{Information entropy ($H\left(A\right)$, $H\left(B\right)$ are the information entropies, $H\left(AB\right)$ the joint entropy\\and $H\left(A|B\right)$ the conditional entropy)}\vspace{0.18cm}\\
		$\mathcal{AB}$ & Entire system or $\mathcal{AB}\equiv\rho_{\mathcal{AB}}$\vspace{0.18cm}\\
		$\mathcal{A}$ & Observed subsystem or $\mathcal{A}\equiv\rho_{\mathcal{A}}$\vspace{0.18cm}\\
		$\mathcal{B}$ & Substance subsystem or $\mathcal{B}\equiv\rho_{\mathcal{B}}$\vspace{0.18cm}\\
		$\mathcal{I}$ & Quantum mutual information\vspace{0.18cm}\\
		$\mathcal{J}$ & Maximum classical correlation\vspace{0.18cm}\\
		$S$ & \tabincell{l}{Von Neumann entropy ($S\left(\mathcal{A}\right)$, $S\left(\mathcal{B}\right)$, $S\left(\mathcal{AB}\right)$, $S\left(\mathcal{A}|\mathcal{B}\right)$ are quantum physics analogies for the\\terms $H\left(A\right)$, $H\left(B\right)$, $H\left(AB\right)$, $H\left(A|B\right)$)}\vspace{0.18cm}\\
		$\mathcal{D}$ & Quantum discord\vspace{0.18cm}\\
		$k$ & Outcome, and $k\in\{0,\ 3\}$\vspace{0.18cm}\\
		$k'$ & Outcome, and $k'\in\{0,\ 31\}$\vspace{0.18cm}\\
		$\{|b_k\rangle\}$ & Von Neumann measurement projective basis on\vspace{0.18cm}\\
		$\Pi_{k}^{\mathcal{A}}$ & Projection operator on subsystem $\mathcal{A}$\vspace{0.18cm}\\
		$\Pi_{k'}^{\mathcal{B}}$ & Projection operator on subsystem $\mathcal{B}$\vspace{0.18cm}\\
		$\rho_k$ & State of the subsystem $\mathcal{B}$ after a measurement of subsystem $\mathcal{A}$ leading to an outcome $k$\vspace{0.18cm}\\
		$\rho_{k'}$ & State of the subsystem $\mathcal{A}$ after a measurement of subsystem $\mathcal{B}$ leading to an outcome $k'$\vspace{0.18cm}\\
		$p_k$ & Probability of outcome $k$\vspace{0.18cm}\\
		$p_{k'}$ & Probability of outcome $k'$\vspace{0.18cm}\\
		$I_{\mathcal{A}}$ & Unit operator\vspace{0.18cm}\\
		$I_{\mathcal{B}}$ & Unit operator\vspace{0.18cm}\\
		$\theta$ & Radian corresponds to the first qubit in observed subsystem\vspace{0.18cm}\\
		$\theta'$& Radian corresponds to the second qubit in observed subsystem\vspace{0.18cm}\\
		$\varphi$ & Radian corresponds to the first qubit in observed subsystem\vspace{0.18cm}\\
		$\varphi'$ & Radian corresponds to the second qubit in observed subsystem\vspace{0.18cm}\\
		$Pr$ & Single-qubit projection operator (e.g. $Pr_0,\ Pr_1,\ Pr_0',\ Pr_1'$)\vspace{0.18cm}\\
		H$_2$ & Hydrogen molecule\\
		$\mathcal{C}$ & Hilbert space\vspace{0.18cm}\\
		$\mathcal{C}'$ & Reduced Hilbert space\vspace{0.18cm}\\
		$|\Psi\rangle$ & Quantum state\vspace{0.18cm}\\
		$|\Psi_{initial}\rangle$ & Initial state\vspace{0.18cm}\\
		$|\Psi_{final}\rangle$ & Final state\vspace{0.18cm}\\
		$\Phi_0$ & Bonding orbital or molecular ground orbital\vspace{0.18cm}\\
		$\Phi_1$ & Antibonding orbital or molecular excited orbital\vspace{0.18cm}\\
		$0_1$ & Excited orbital in one hydrogen atom\vspace{0.18cm}\\
		$0_2$ & Excited orbital in another hydrogen atom\vspace{0.18cm}\\
		$\uparrow$ & Spin up\vspace{0.18cm}\\
		$\downarrow$ & Spin down\vspace{0.18cm}\\
		$at$ & Atom\vspace{0.18cm}\\
		$pht$ & Photon\vspace{0.18cm}\\
		$cb$ & Covalent bond\vspace{0.18cm}\\
		$n$ & Nucleus\vspace{0.18cm}\\
		$\Omega$ & Transition frequency for electron in molecule\vspace{0.18cm}\\
		$\Omega^{\uparrow}$ & Transition frequency for electron with $\uparrow$ in molecule\vspace{0.18cm}\\
		$\Omega^{\downarrow}$ & Transition frequency for electron with $\downarrow$ in molecule\vspace{0.18cm}\\
		$\Omega_{at}^{\uparrow}$ & Transition frequency for electron with $\uparrow$ in atom\vspace{0.18cm}\\
		$\Omega_{at}^{\downarrow}$ & Transition frequency for electron with $\downarrow$ in atom\vspace{0.18cm}\\
		$\omega$ & Phononic mode\vspace{0.18cm}\\
		$H_{sys}$ & Hamiltonian\vspace{0.18cm}\\
		$H_{sys}'$ & Reduced Hamiltonian\vspace{0.18cm}\\
		$h$ & Planck constant\vspace{0.18cm}\\
		$\hbar$ & Reduced Planck constant or Dirac constant, and $\hbar=h/2\pi$\vspace{0.18cm}\\
		$\eta$ & Maximum ratio of coupling strength to frequency\vspace{0.18cm}\\
		$g$ & Coupling strength of photon and the electron (e.g. $g_{\Omega^{\uparrow}},\ g_{\Omega^{\downarrow}},\ g_{\omega}$)\vspace{0.18cm}\\
		$\zeta$ & Nucleus tunnelling strength or atom leap strength\vspace{0.18cm}\\
		$\omega_c$ & Cavity frequency\vspace{0.18cm}\\
		$\omega_a$ & Transition frequency\vspace{0.18cm}\\
		$a$ & \tabincell{l}{Photon annihilation operator (e.g. $a_{\Omega^{\uparrow}},\ a_{\Omega^{\downarrow}},\ a_{\omega}$), and its hermitian conjugate\\operator --- $a^{\dag}$}\vspace{0.18cm}\\
		$\sigma$ & \tabincell{l}{Interaction operator of atom with the electromagnetic field of the cavity (e.g. $\sigma_{\Omega^{\uparrow}},\ \sigma_{\Omega^{\downarrow}},\ \sigma_{\omega}$,\\$\sigma_n$), and its hermitian conjugate operator --- $\sigma^{\dag}$}\vspace{0.18cm}\\
		$\rho$ & Density matrix (e.g. $\rho_{\mathcal{AB}},\ \rho_{\mathcal{A}},\ \rho_{\mathcal{B}}$, etc)\vspace{0.18cm}\\
		$\mathcal{L}\left(\rho\right)$ & Lindblad superoperator\vspace{0.18cm}\\
		$\mathcal{K}$ & Graph of the potential photon dissipations between the states that are permitted\vspace{0.18cm}\\
		$\mathcal{K}'$ & Graph of the potential photon influxes between the states that are permitted\vspace{0.18cm}\\
		$L_k\left(\rho\right)$ & Standard dissipation superoperator\vspace{0.18cm}\\
		$L_{k'}\left(\rho\right)$ & Standard influx superoperator\vspace{0.18cm}\\
		$\gamma_{k}$ & Total spontaneous emission rate for photon\vspace{0.18cm}\\
		$\gamma_{k'}$ & Total spontaneous influx rate for photon\vspace{0.18cm}\\
		$A_k$ & Lindblad or jump operator of system, and its hermitian conjugate operator --- $A_k^{\dag}$\vspace{0.18cm}\\
		$c$ & Normalization factors (e.g. $c,\ c',\ c'',\ c'''$)\vspace{0.18cm}\\
		$C$ & Constants (e.g. $C_0,\ C_1,\ C_2,\ C_3,\ C_4$)\vspace{0.18cm}\\
		$N$ & Number of particles, where $\sum_{\alpha}n_{\alpha}=N$
	\end{longtable}
	
	\twocolumngrid

\bibliography{bibliography}
	
\end{document}